\documentclass[twocolumn,prc,aps,showpacs,amsmath,amssymb]{revtex4}
\usepackage{graphicx}

 \def\gsim{\mathrel{\rlap{\lower4pt\hbox{\hskip1pt$\sim$}}
 \raise1pt\hbox{$>$}}}

 \newcommand\la{\langle}
 \newcommand\ra{\rangle}
 \newcommand\beq{\begin{equation}}
 
 \newcommand\eeq{\end{equation}}
 \newcommand\beqn{\begin{eqnarray}}
 \newcommand\eeqn{\end{eqnarray}}

\def\lsim{\mathrel{\rlap{\lower4pt\hbox{\hskip1pt$\sim$}}
 \raise1pt\hbox{$<$}}}
 \def\gsim{\mathrel{\rlap{\lower4pt\hbox{\hskip1pt$\sim$}}
 \raise1pt\hbox{$>$}}}
\def\Re{\,\mbox{Re}\,}

\def\fm{\,\mbox{fm}}
\def\GeV{\,\mbox{GeV}}

\def\sq{\sigma_{\bar qq}}
\def\sg{\sigma_{gg}}

\def\verq{\vec r^{\,q}}

\def\verqaa{\vec r_1^{\,q_1}}

\def\verqba{\vec r_2^{\,q_1}}
\def\verqbb{\vec r_2^{\,\bar q_2}}
\def\verqbc{\vec r_2^{\,q_3}}
\def\verqca{\vec r_3^{\,q_1}}
\def\verqcb{\vec r_3^{\,\bar q_2}}
\def\verqcc{\vec r_3^{\,q_3}}
\def\verqda{\vec r_4^{\,q_1}}
\def\verqdb{\vec r_4^{\,\bar q_2}}
\def\verqdc{\vec r_4^{\,q_3}}
\def\verga{\vec r_1^{\,g}}
\def\vergb{\vec r_2^{\,g}}

\def\vesq{\vec s^{\,q}}
\def\vesqaa{\vec s_1^{\,q_1}}

\def\vesqba{\vec s_2^{\,q_1}}

\def\vesqca{\vec s_3^{\,q_1}}
\def\vesqcb{\vec s_3^{\,\bar q_2}}
\def\vesqcc{\vec s_3^{\,q_3}}
\def\vesqda{\vec s_4^{\,q_1}}
\def\vesqdb{\vec s_4^{\,\bar q_2}}
\def\vesqdc{\vec s_4^{\,q_3}}
\def\vesga{\vec s_1^{\,g}}
\def\vesgb{\vec s_2^{\,g}}
\def\vesgc{\vec s_3^{\,g}}

\def\verqa{\vec r^{\,q_1}}
\def\verqb{\vec r^{\,\bar q_2}}
\def\verqc{\vec r^{\,q_3}}
\def\verg{\vec r^{\,g}}

\def\vesqa{\vec s^{\,q_1}}
\def\vesqb{\vec s^{\,\bar q_2}}
\def\vesqc{\vec s^{\,q_3}}
\def\vesg{\vec s^{\,g}}

\def\vR{\{\vec R\}}

\begin{document}

\title{\bf Quantum-mechanical description of in-medium fragmentation}

\author{B.Z.~Kopeliovich$^{1-3}$}
\author{H.-J.~Pirner$^{2}$}
\author{I.K.~Potashnikova$^1$}
\author{Ivan~Schmidt$^1$}
\author{A.V.~Tarasov$^{2,3}$}
\author{O.O.~Voskresenskaya$^{3}$}

\affiliation{$^1$Departamento de F\'\i sica
y Centro de Estudios
Subat\'omicos,\\ Universidad T\'ecnica
Federico Santa Mar\'\i a, Casilla 110-V, Valpara\'\i so, Chile\\
{$^2$Institut f\"ur Theoretische Physik der Universit\"at,
Philosophenweg 19, 69120
Heidelberg, Germany}\\
{$^3$Joint Institute for Nuclear Research, Dubna, Russia}
}


\begin{abstract}
We present a quantum-mechanical description of quark-hadron
fragmentation in a nuclear environment. It employs the path-integral
formulation of quantum mechanics, which takes care of all phases and
interferences, and which contains all relevant time scales, like
production, coherence, formation, etc. The cross section includes
the probability of pre-hadron (colorless dipole) production both
inside and outside the medium. Moreover, it also includes
inside-outside production, which is a typical quantum-mechanical
interference effect (like twin-slit electron propagation). We
observe a substantial suppression caused by the medium, even if the
pre-hadron is produced outside the medium and no energy loss is
involved. This important source of suppression is missed in the
usual energy-loss scenario interpreting the effect of jet quenching
observed in heavy ion collisions. This may be one of the reasons of
a too large gluon density, reported by such analyzes.

 \end{abstract}


\maketitle

\section{Introduction}

\begin{figure}[htb]
 \includegraphics[width=8cm]{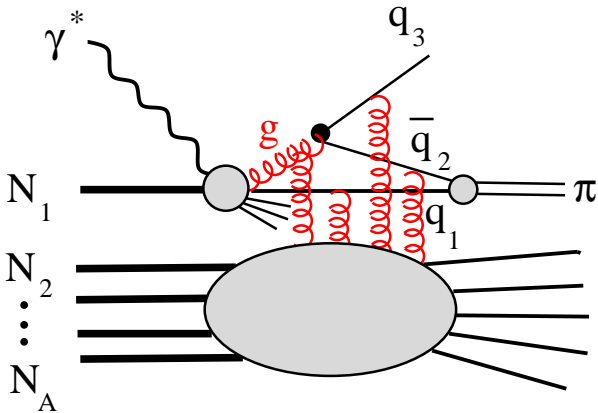}
 \caption{\label{gamma-a} Reaction $\gamma^*p\to \pi X$. The incoming virtual
  proton is absorbed by a valence quark of a bound nucleon, leading to the production
  of a quark and gluon, $\gamma^*q\to q_1g$. The gluon decays to a $\bar q_2q_3$, and
  $q_1-\bar q_2$ fuse to the final pion. The intermediate and final partons can
  experience multiple interactions in the nuclear medium.}
 \end{figure}

Hadronization in a nuclear environment has always been a precious
source of information about the space time pattern of hadronization.
This process is characterized by the production length of a
pre-hadron (a colorless dipole), with the subsequent development of
the hadronic wave function. A perturbative description of this process as radiation of a $\bar qq$ pair and creation of colorless dipole evolving to the final pion, as is illustrated in Fig.~\ref{gamma-a},
was proposed in \cite{knp,knph}. 
Although the model was partly probabilistic, it led to rather successful description \cite{knph} and even prediction \cite{knp} for
semi-inclusive hadron production in deep-inelastic scattering (DIS)
\cite{hermes}. A pure classical string model \cite{k,pir1,pir2} fitted to data also 
leads to a good agreement. No fully quantum mechanical description
of in-medium hadronization has been available  so far.

The widely debated question whether the hadronization process ends
by a leading  pre-hadron production "within or without the medium?"
\cite{knph}, strictly speaking does not have a definite answer. In
quantum mechanics a pre-hadron may be created both inside and
outside the medium, and interference of the corresponding amplitudes
is important, as is demonstrated below.

Here we are developing the model \cite{knp, knph} of perturbative
hadronization and employ the Berger model \cite{berger}, which  we improved in a recent paper \cite{born}. In this model an energetic quark produced in a hard reaction, like $e^+e^-$ annihilation, DIS, or high-$p_T$ scattering, creates a
leading pion, which carries a major fraction $z_h\to 1$ of the quark
momentum, via perturbative radiation of a gluon decaying into a
$\bar qq$ pair.  Then the $\bar q$ fuses with the original quark
into the pion, as is illustrated in Fig.~\ref{gamma-a} for
$\gamma^*$-nucleus collision.
In general, fragmentation is process dependent, due to the higher
twist  terms calculated in \cite{berger,born}, which we neglect
here. Therefore our results are applicable to any hard reaction
leading to the production of a quark jet.

The main approximations in the Berger model \cite{berger} are: (i)
the calculation is done to lowest order of $(1-z_h)$, which is
considered as a small parameter; (ii) the pion is treated as a $\bar
qq$ state with no transverse or longitudinal motion of the quarks.
New calculations without these approximations were done in
\cite{born}, where the cross section of the process $\gamma^*p\to\pi
X$ was calculated to all orders of $(1-z_h)$ and with a realistic
model for the light-cone pion wave function

The calculations performed in \cite{berger,born} were done in
momentum representation, employing the Feynman diagram technique,
and for the case of a proton target. Usually for nuclear targets the
impact parameter representation is more effective, since at high
energies the impact parameters do not vary during propagation
through the nucleus, which allows to apply a Glauber-like
eikonalization. If the energy is not sufficiently high to freeze the
impact parameters, one should integrate over all possible paths of
the propagating partons.

 Here we
employ the light cone Greens function formalism which is the essential tool for the 
calculation of the nuclear effects \cite{kz91,rauf,kst1, kst2}. We calculate the ratio of
$p_T$-integrated cross sections, 
\beq
R_{A/p}(z_h)=\frac{d\sigma(\gamma^*A\to\pi X)/dz_h}
{A\,d\sigma(\gamma^*N\to\pi X)/dz_h}, 
\label{50} 
\eeq 
as function of pion fractional light-cone momentum $z_h$. We consider gluon
decay and pre-hadron ($\bar q_2q_1$) production both inside and outside
the nucleus. However, these two possibilities can be clearly separated only
in a probabilistic approach. In fact, in quantum mechanics the cross
section is related to the square of the process amplitude, and the
production of the pre-hadron occurs in the direct and conjugated
amplitudes at different points. This also implies the appearance of
an interference term between inside-outside production, which we
consider in our calculation.

The paper is organized as follows. In section~\ref{amplitude} we
introduce the kinematic variables and present the general structure
of the amplitude of jet production. The amplitude is written in
impact parameter presentation, which is especially convenient for
the calculation of multiple interaction effects.

In section~\ref{green} the cross section of inclusive hadron
production is expressed via the light-cone Green functions
describing the propagation of parton ensembles through a nuclear
medium. The composition of these ensembles correlates with the
coordinates of gluon-to-$\bar qq$ decays in the two amplitudes,
direct and complex conjugated. Different colorless parton ensembles
propagating through the medium interact with different multi-parton
cross sections. These cross sections are derived in \ref{sigmas} and
expressed in terms of phenomenologically known cross sections of
$\bar qq$ dipoles interacting with a proton. Using known
multi-parton cross sections one can solve the light-cone
Schr\"odinger equations for the Green function describing
propagation of partonic ensembles. These solutions are found in
\ref{W}.

In section~\ref{3parts} the cross section is presented as a sum of
three terms, Eq.~(\ref{2600}), the first two terms corresponding to
the amplitudes for gluon decay both inside or outside the nucleus,
and the third term corresponding to inside-outside interference.
These three terms are further evaluated employing different models for the 
pion wave function, which differ in the assumptions about the
longitudinal and transverse momentum distributions of valence quarks
in the pion. The unrealistic assumption of the Berger model that the
quarks have no Fermi motion results in a complete absence of nuclear
effects, while more realistic models lead to a considerable nuclear
suppression. 

Section~\ref{results} presents the main results and observations of this paper, as well
as an outlook to future developments.\\

\section{The process \boldmath$\gamma^*p\to\pi X$ in impact parameter representation}\label{amplitude}

\subsection{The amplitude}

In what follows we consider the reaction $\gamma^*p\to \pi X$ as an
example of a hard reaction, neglecting the higher-twist terms
calculated in \cite{born}. Therefore, the space time development of
this process and all the results are valid for any hard process
producing a quark jet.

The hard reaction  $\gamma^*A\to \pi X$ can be considered as a
three-step process, as is illustrated in Fig.~\ref{gamma-a}. In the
first stage the incoming virtual photon knocks out  a quark and a
gluon of a bound nucleon, $\gamma^*+p\to q_1+g$, which carry
practically all the energy of the photon.
 We assume that Bjorken $x$ is sufficiently large to neglect the contribution of the sea.
 It also allows to treat the gluon radiation process as incoherent. Strictly speaking one should integrate over the longitudinal coordinate of gluon radiation from $-\infty$ \cite{kst2}, however
 only a part of this path, $\Delta z\sim 1/x m_N$, contributes coherently. We neglect $\Delta z$ assuming that $x$ is large. Thus, in the light-cone approach one can consider gluon radiation as instantaneous from the point of hard interaction, although the quark-gluon pair loose coherence at longer distances. One can come to the same conclusion analyzing the Feynman graphs. The corresponding space-time structure of DIS is studied in \ref{avt}.
 
The second stage is the dissociation of the radiated gluon into a
quark pair, $g\to \bar q_2+q_3$, and the propagation of the
colorless dipole $\bar q_2q_1$ through the nucleus.

The third stage is the projection  of the colorless dipole into a
pion, $\bar q_2+q_1\to\pi$. The gluon and quarks propagating through
the nucleus are assumed to experience only soft final state
interactions (soft gluonic exchanges) with other
nucleons-spectators, which usually cause attenuation.

 The amplitude of this process can be represented as,
 \begin{widetext}
   \beqn
  M&=& \int \limits_0^1d\alpha \int d^2\kappa
  \lim\limits_{z_3\to\infty}
  \int\limits_{z_1}^{z_3} dz_2\,e^{i(\Delta+io)z_2}\,
  \Phi_\pi(\alpha,\vec\kappa)
    \nonumber\\ &\times&
 \int d^2r_1^g d^2r_1^{q_1}d^2r_2^g\,d^2r_3^{q_1} d^2r_3^{\bar q_2} d^2 r_3^{q_3}\,\exp\left[-i\vec p_1\vec r_3^{\,q_1}
-i\vec p_2\vec r_3^{\,\bar q_2}-i\vec p_3\vec r_3^{\,q_3}\right]
\Gamma(\vec r_1^{\,g},\vec r_1^{\,q_1})\,
G_g(z_2,z_1;\vec r_2^{\,g}\vec r_1^{\,g};E_g;\{\vec R\})
 \nonumber\\ &\times&
G_{q_1}(z_3,z_1;\vec r_3^{\,q_1}\vec r_1^{\,q_1};E_1;\{\vec R\})\,
G_{\bar q_2}(z_3,z_2;\vec r_3^{\,\bar q_2}\vec r_2^{\,g};E_2;\{\vec R\})\,
G_{q_3}(z_3,z_2;\vec r_3^{\,q_3}\vec r_2^{\,g};E_3;\{\vec R\})
\label{100}
 \eeqn
 \end{widetext}
Here $\Phi_\pi$ is the wave function of the $\bar qq$ Fock component of the produced pion;
\beqn
\vec p_1&=&\alpha\vec p_\pi+\vec\kappa;
 \nonumber\\
 \vec p_2 &=&(1-\alpha)\vec p_\pi-\vec\kappa;
 \label{200}
  \eeqn
$\vec p_\pi$ is the pion transverse momentum;
$\vec\kappa=(1-\alpha)\vec p_1-\alpha\vec p_2$ is the relative
transverse momentum of  the $q_1$ and $\bar q_2$ in the
 pion;
 \beqn
\alpha&=&\frac{E_1}{E_\pi};
 \nonumber\\
 \Delta&=&\frac{m_q^2}{2E_2}+\frac{m_q^2}{2E_3};
 \label{300}
  \eeqn
 and $E_1,\ E_2,\ E_3$ and $E_g$ are the energies of the three quarks and gluon.

The longitudinal coordinates $z_i$ are defined as follows. $z_1$ is
the coordinate of the collision between the virtual photon and the
nucleon; $z_2$ is the longitudinal coordinate of the point of
dissociation $g\to q_3\bar q_2$; $\Gamma(\vec r_1^{\,g},\vec
r_1^{\,q_1})$ is the amplitude of the process $\gamma^*N\to q_1 g
X_1$, with the original impact parameters of the produced quark
($\vec r_1^{\,q_1}$) and gluon ($\vec r_1^{\,g}$), at the point with
coordinate $z_1$.

The propagation functions (Green's functions) $G(z_f,z_{in};\vec
r_f,\vec r_{in};E;\{\vec R\})$ in Eq.~(\ref{100}) describing the
propagation of the fast quarks $q_1,\ \bar q_2,\ q_3$ and the gluon
in the medium, will be derived in Sect.~\ref{green}. 
Besides the initial ($z_{in},\vec r_{in}$) and final
($z_f,\vec r_f$) positions, they also depend on the coordinates
$\{\vec R\}$ of the spectator nucleons with whom they interact via
soft gluonic exchanges. 

 The cross section of pion production off a nucleus is given by the amplitude
 squared and averaged over the positions of all nucleons in the nucleus,
 \beq
 \frac{d\sigma}{d^2p_3 d^2p_\pi dz_h} =
 A\int d^2b\,dz_1\,\rho_A(b, z_1)\,
 \left\la|M|^2\right\ra_{\{\vec R\}}.
 \label{500}
 \eeq
Here $z_h$ is the fraction of the photon light-cone momentum carried
by the pion.  We singled out the integration over the coordinates of
the "active" nucleon participating in the hard collision with the
virtual photon.

 \subsection{\boldmath$Q^2$-dependence}\label{q2-dep}

 At large photon virtuality $Q^2$ the relative quark-gluon separation is small,
 $\left|\vec r_1^{\,g}- \vec r_1^{\,q_1}\right|\sim 1/Q$,
 and the nuclear effects become independent
 of $Q^2$. Indeed,  according to the uncertainty principle the smaller is the quark-gluon separation $r$, the faster they are expanding with transverse momentum $k_T\sim1/r$,
 \beq
 \frac{dr}{dt}=\frac{k_T}{E}\approx \frac{1}{r\,E}.
 \label{350}
 \eeq
 Here $\vec r=\verqa-\verg$; $E=E_1+E_g$ is the total energy of the jet.

 If the initial size is small, then after a while its smallness will be forgotten. Indeed,
 the solution of Eq.~(\ref{350}) reads,
 \beq
 r^2(t)=\frac{2t}{E}+{1\over Q^2}.
 \label{370}
 \eeq
 At sufficiently long time intervals,
 \beq
 t\gg \frac{1}{4m_N\,x_{Bj}},
 \label{380}
 \eeq
where $x_{Bj}= Q^2/2m_NE$ is the Bjorken scaling
variable. At large $Q^2$ the second term in  (\ref{370}) can be
neglected, and therefore transverse size of the quark-gluon pair
does not depend on $Q^2$ any more. If this time interval is
significantly shorter than the mean free path of  partons in the
medium, no $Q^2$ dependence of nuclear effects should be expected.
In cold nuclear matter the typical mean free path is several Fermi,
so for $t\sim 1\fm$ we expect a very weak $Q^2$ dependence when
$x_{Bj}\gg 0.05$. This condition is well satisfied in the region of
$x_{Bj}\gsim0.1$ dominated by valence quarks, in which we are
focused. This effect probably explains the very weak dependence on
$Q^2$ of nuclear ratios observed in the HERMES experiment
\cite{q-dep-herm}.

In what follows we assume that $x_{Bj}$ is sufficiently large to
neglect the second term in (\ref{370}), which is equivalent to the
approximation in (\ref{100}),
 \beq
 \Gamma(\vec r_1^{\,g},\vec r_1^{\,q_1})\approx
 \tilde\Gamma\left( \vec r_1^{\,g}\right)\,
 \delta\left( \vec r_1^{\,g}-\vec r_1^{\,q_1}\right) .
 \label{400}
 \eeq

 \section{Green function formalism for propagation of partons in a medium}\label{green}

Although the hard reaction (DIS in Fig.~\ref{gamma-a}) occurs on different nucleons incoherently, the multiple final state interactions of the produced partons proceed further coherently. Indeed, the mean transverse momentum squared gained by a quark propagating through a heavy nucleus, as measured in the Drell-Yan reaction at $800\GeV$, is very small of the order of $\Delta p_T^2\sim0.1-0.2\GeV^2$ \cite{e866}, and is even several time smaller at lower energies \cite{hermes-pt,jlab}. So a quark of energy $E_q\sim10\GeV$ interacts with coherence length $l_c=2E_q/\Delta p_T^2\sim40\fm$, which is quite long compared to the nuclear size.

\begin{figure}[htb]
 \includegraphics[width=8cm]{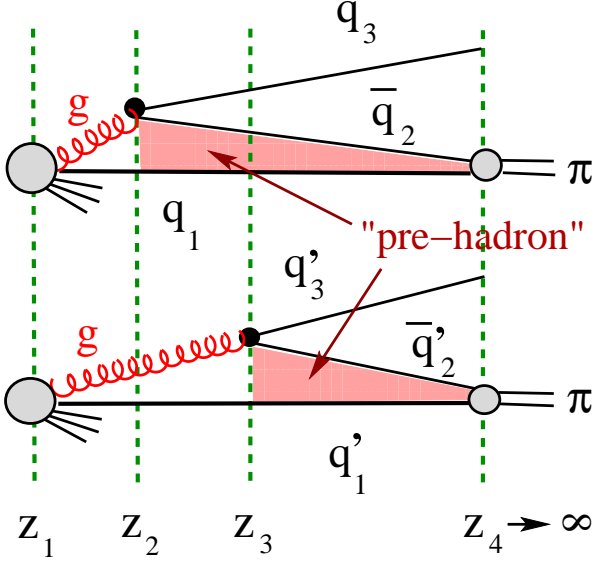}
 \caption{\label{2amplitudes} Graphical representation for the direct and conjugated
 amplitudes of the process $\gamma^*p\to \pi X$. The DIS hard process occurs incoherently
 on the same nucleon with longitudinal coordinate $z_1$. The radiated gluon decays to
 $\bar q_2 q_3$ coherently in the two amplitudes at the points $z_2$ and $z_3$ respectively.
 The colorless pre-hadrons (dipoles) $\bar q_2 q_1$ and $\bar q_2^{\,\prime} q_1^\prime$
 created at $z_2$ and $z_3$ respectively, are projected to the pion wave function in each
 of the two amplitudes.
  }
 \end{figure}

The cross section for the reaction  $\gamma^*p\to \pi X$ corresponds
to the product of the direct and conjugated amplitudes, presented
graphically in Fig.~\ref{2amplitudes}.
The two amplitudes correspond to different impact parameters, ($\vec
r_i$) and ($\vec s_i$), of the participating partons, and different
longitudinal coordinates, $z_2$ and $z_3$, of $g\to\bar qq$ decay.

One can see that specific partonic ensembles propagate through
different intervals of the longitudinal coordinate. It is $\{2g\bar
q_1q_1\}$ in the interval $z_1-z_2$, $\{gq_3\bar q_2q_1\bar q_1\}$
in the interval $z_2-z_3$, $\{\bar q_1q_1\bar q_2q_2\bar q_3q_3\}$
in the interval $z_3-z_4$.

In the case of free propagation in vacuum the Green functions
introduced in (\ref{100}) have a simple form,
  \beqn
  G(z_2,z_1;\vec r_2,\vec r_1;E)&=&
  -\frac{iE}{2\pi(z_2-z_1)}\,
  \Theta(z_2-z_1)
\nonumber\\ &\times&
  \exp\left[\frac{iE(\vec r_2-\vec r_1)^2}{2(z_2-z_1)}\right]\,
 \label{600}
  \eeqn

In equation (\ref{500}) the amplitude (\ref{100}) squared contains
bilinear combinations of Green functions with the same initial and
final longitudinal coordinates, but different impact parameters. For
instance the product $G_g(z_2,z_1;\vec r_2^{\,g}\vec
r_1^{\,g};E_g;\{\vec R\}) G^*_g(z_2,z_1;\vec s_2^{\,g}\vec
s_1^{\,g};E_g;\{\vec R\})$. In the case of in-medium interacting
partons, the Feynman path-integral representation is more
appropriate. It allows to introduce differential equations for
products of the Green functions and their conjugated functions,
averaged over nucleon coordinates in the nucleus (see in more detail
in the Appendix of \cite{kst1}). Such a combination, describing the
propagation of the gluon and quark $q_1$ between points $z_1$ and
$z_2$, has the form,
  \beqn
  &&W_1=\biggl\la
  G_g(z_2,z_1;\vec r_2^{\,g}\vec r_1^{\,g};E_g;\{\vec R\})
 \nonumber\\ &\times&
G_{q_1}(z_2,z_1;\vec r_2^{\,q_1}\vec r_1^{\,q_1};E_1;\{\vec R\})\,
 G^*_g(z_2,z_1;\vec s_2^{\,g}\vec s_1^{\,g};E_g;\{\vec R\})
 \nonumber\\ &\times&
G^*_{q_1}(z_2,z_1;\vec s_2^{\,q_1}\vec s_1^{\,q_1};E_1;\{\vec R\})
\biggr\ra_{\{\vec R\}},
\label{700}
\eeqn
which is a part of the final equation (\ref{1300}) (see below). It satisfies the equations,
\beqn
&& i\,\frac{\partial W_1}{\partial z_2}=
\biggl[-\frac{\Delta r_2^{q_1}}{2 E_g} -
\frac{\Delta r_2^g}{2E_g}+
\frac{\Delta s_2^{q_1}}{2\bar E_g} +
\frac{\Delta s_2^g}{2\bar E_g}
\nonumber\\ &-&
{i\over2}\,\rho_A(b,z_2)\,\Sigma_1(\vec r_2^{\,q_1},\vec s_2^{\,g},\vec s_2^{\,q_1},\vec r_2^{\,g})\biggr]W_1;
\label{800}
\eeqn

\beqn
&&i\,\frac{\partial W_1}{\partial z_1}=
\biggl[-\frac{\Delta r_1^{q_1}}{2 E_g} -
\frac{\Delta r_1^g}{2E_g}+
\frac{\Delta s_1^{q_1}}{2\bar E_g} +
\frac{\Delta s_1^g}{2\bar E_g}
\nonumber\\ &-&
{i\over2}\,\rho_A(b,z_1)\,\Sigma_1(\vec r_1^{\,q_1},\vec s_1^{\,g},\vec s_1^{\,q_1},\vec r_1^{\,g})\biggr]W_1,
\label{850}
\eeqn
with initial conditions,
\beqn
W_1\bigr|_{z_2<z_1} &=& 0
\nonumber\\
W_1\bigr|_{z_2=z_1} &=& \delta(\vec r_2^{\,g}-\vec r_1^{\,g})\,
\delta(\vec r_2^{\,q_1}-\vec r_1^{\,q_1})
\nonumber\\ &\times&
\delta(\vec s_2^{\,g}-\vec s_1^{\,g})\,
\delta(\vec s_2^{\,q_1}-\vec s_1^{\,q_1}).
\label{900}
\eeqn

Here $\vec b$ is the impact parameter of the virtual photon, and the
nuclear density is normalized to one, $\int d^2b dz\,\rho_A(b,z) =
1$. $\Sigma_1$ is the total cross section of a 4-parton colorless
system $gq_1\,\bar g\bar q_1$ interacting with a nucleon target. It
is important to notice that $gq_1$ and $\bar g\bar q_1$ are in color
triplet and anti-triplet states respectively, while $\bar gg$ and
$\bar q_1q_1$ are color singlets.

The 4-body cross section $\Sigma_1$ can be represented as a linear
superposition of elementary dipole cross sections of interaction of
a colorless $\bar qq$ dipole with a nucleon, for which there exists
a well developed phenomenology. A derivation presented in
\ref{sigmas} results in the expression,
\beqn
&&\Sigma_1(\vec
r^{\,q_1},\vec s^{\,g},\vec s^{\,q_1},\vec r^{\,g})=
{9\over8}\,\biggl[ \sigma_{\bar qq}(\vec r^{\,g}-\vec r^{\,q_1})
\nonumber\\ &+& \sigma_{\bar qq}(\vec s^{\,g}-\vec s^{\,q_1})-
\sigma_{\bar qq}(\vec r^{\,g}-\vec s^{\,q_1})- \sigma_{\bar qq}(\vec
s^{\,g}-\vec r^{\,q_1})\biggr] \nonumber\\ &+& \sigma_{\bar qq}(\vec
r^{\,q_1}-\vec s^{\,q_1})+ {9\over4}\sigma_{\bar qq}(\vec
r^{\,g}-\vec s^{\,g}) \label{1000}
\eeqn

The solution of Eq.~(\ref{800}) with a realistic dipole cross
section, and which is valid for large $\bar qq$ separations, can
usually be obtained only numerically (see \cite{jan}). However, in
the reaction under consideration, with a highly virtual photon, the
typical separations are small and one may rely on the approximation
\cite{zkl},
 \beq
\sigma_{\bar qq}(\vec r_1-\vec r_2)=C\,(\vec r_1-\vec r_2)^2,
\label{1100}
 \eeq
where the factor $C$ is known from phenomenology. It depends on
energy and should be properly chosen depending on the energy of the
photon. With this approximation and a constant nuclear density,
$\rho_A(b,z)=\rho_0\,\Theta(R_A^2-b^2-z^2)$, the equations
Eq.~(\ref{800}) can be solved analytically (see \ref{sigma4}).

To progress further, for propagation of the quark $q_1$ we employ the relation,
 \beqn
 G(z_3,z_1;\vec r_3,\vec r_1;E;\{\vec R\})&=&
 \int d^2 r_2\,
 G(z_3,z_2;\vec r_3,\vec r_2;E;\{\vec R\})\
 \nonumber\\ &\times&
G(z_2,z_1;\vec r_2,\vec r_1;E;\{\vec R\}),
\label{1200}
 \eeqn
 which assumes that $z_1<z_2<z_3$. Then the cross section Eq.~(\ref{500}) can be represented as,
 \begin{widetext}
\beqn
&& \frac{d\sigma}{d^2p_3 d^2p_\pi dz_h} =
 A\,\frac{F_2(x)}{x}
 \lim\limits_{z_4\to\infty} 2\Re
 \int d^2b\int dz_1 dz_2 dz_3\,
 \rho_A(b,z_1)\Theta(z_3-z_2)\Theta(z_2-z_1)\,
 e^{i(\Delta+io)z_2-i(\Delta-io)z_3}
 \nonumber\\ &\times&
 \int d^2r_1^g d^2r_1^{q_1} d^2s_1^g d^2s_1^{q_1}\,
 \Gamma(\vec r_1^{\,g},\vec r_1^{\,q_1})
  \nonumber\\ &\times&
 \int d^2r_1^g d^2r_2^{q_1} d^2s_2^g d^2s_2^{q_1}\,
 W_1(z_2,z_1;\vec r_2^{\,g},\vec r_2^{\,q_1};
 \vec s_2^{\,g},\vec s_2^{\,q_1};
 \vec r_1^{\,g},\vec r_1^{\,q_1};
 \vec s_1^{\,g},\vec s_1^{\,q_1};
 E_g,\bar E_g,E_1,\bar E_1)
 \nonumber\\ &\times&
 \int d^2r_3^{q_3} d^2r_3^{\bar q_2} d^2r_3^{\bar q_1} d^2s_3^g d^2s_3^{q_1}\,
 W_2(z_3,z_2;\vec r_3^{\,q_3},\vec r_3^{\,\bar q_2},\vec r_3^{\,q_1};
 \vec s_3^{\,g},\vec s_3^{\,q_1};
 \vec r_2^{\,g},\vec r_2^{\,q_1};
 \vec s_2^{\,g},\vec s_2^{\,q_1};
 E_1,E_2,E_3,\bar E_g,\bar E_1)
 \nonumber\\ &\times&
  \int d^2r_4^{q_3} d^2r_4^{\bar q_2} d^2r_4^{\bar q_1} d^2s_4^{q_3}
  d^2s_4^{\bar q_2} d^2s_3^{q_1}\,
 \nonumber\\ &\times&
W_3(z_4,z_3;\vec r_4^{\,q_3},\vec r_4^{\,\bar q_2},\vec r_4^{\,q_1};
 \vec s_4^{\,q_3}, \vec s_4^{\,\bar q_2}, \vec s_3^{\,q_1};
 \vec r_3^{\,q_3},\vec r_3^{\,\bar q_2},\vec r_3^{\,q_1};
 \vec s_3^{\,g},\vec s_3^{\,\bar q_2},\vec s_3^{\,q_1};
 E_1,E_2,E_3,\bar E_1,\bar E_2,\bar E_3)
 \nonumber\\ &\times&
\int d^2\kappa\, d^2\kappa' d\alpha \,d\bar\alpha\,
\Phi_\pi(\alpha,\vec\kappa)\,\Phi_\pi^*(\bar\alpha,\vec\kappa^{\,\prime})
 \exp\bigl[i\vec p_1 ^{\,\prime}\cdot\vec s_4^{\,q_1}
+ i\vec p_2 ^{\,\prime}\cdot\vec s_4^{\,\bar q_2}
 + i\vec p_3 ^{\,\prime}\cdot\vec s_4^{\,q_3} -
 i\vec p_1 ^{\,\prime}\cdot\vec r_4^{\,q_1}-
 i\vec p_2 ^{\,\prime}\cdot\vec r_4^{\,\bar q_2} -
 i\vec p_3 ^{\,\prime}\cdot\vec r_4^{\,q_3} \bigr]
  \label{1300}
 \eeqn
\end{widetext}

Here $\vec r^{\,q_i}_j$ and $\vec s^{\,q_i}_j$ are the transverse
coordinates of the quark $q_i$ ($i=1,2,3$) at the point with
longitudinal coordinate $z_j$ ($j=1,2,3,4$) for the direct and
conjugated amplitudes respectively (same for the gluons); $\alpha$
and $\bar\alpha$ are the fractional momenta of the quark $q_1$
within the pion in the direct and conjugated amplitudes
respectively. Correspondingly, the transverse momenta of the quarks
$q_1$ and $\bar q_2$ are,
\beqn \vec p_1&=&\alpha\vec
p_\pi+\vec\kappa;\ \ \ \ \ \ \ \ \ \ \ \vec
p_1^{\,\prime}=\bar\alpha\vec p_\pi+\vec\kappa^\prime;
\nonumber\\
\vec p_2&=&(1-\alpha)\vec p_\pi-\vec\kappa;\ \ \ \
\vec p_2^{\,\prime}=(1-\bar\alpha)\vec p_\pi-\vec\kappa^\prime;
\nonumber\\
\vec p_3^{\,\prime}&=&\vec p_3;
\label{1400}
\eeqn
The energies of the participating quarks and gluon in the two amplitudes read,
\beqn
E_1&=&\alpha z_h E;
\nonumber\\
E_2&=&(1-\alpha)z_h E;
\nonumber\\
E_g&=&(1-\alpha z_h)E;
\nonumber\\
\bar E_1&=&\bar\alpha z_h E;
\nonumber\\
\bar E_2&=&(1-\bar\alpha)z_h E;
\nonumber\\
\bar E_g&=&(1-\bar\alpha z_h)E;
\nonumber\\
\bar E_3&=&E_3=(1-z_h)E.
\label{1500}
 \eeqn

The functions $W_2$ and $W_3$ in (\ref{1300}) describing the
propagation of partonic ensembles through the intervals $z_2-z_3$
and $z_3-z_4$ respectively, are defined similar to Eq.~(\ref{700}),
\begin{widetext}
\beqn
W_2&=&\biggl\la
G_{q_3}\bigl[z_3,z_2;\verqcc,\verqbc;E_3;\vR\bigr]\,
G_{\bar q_2}\bigl[z_3,z_2;\verqcb,\verqbb;E_2;\vR\bigr]\,
G_{q_1}\bigl[z_3,z_2;\verqca,\verqba;E_1;\vR\bigr]
\nonumber\\&\times&
G^*_{q_1}\bigl[z_3,z_2;\vesqca,\vesqba;\bar E_1;\vR\bigr]\,
G^*_{g}\bigl[z_3,z_2;\vesgc,\vesgb;\bar E_g;\vR\bigr]
\biggr\ra_{\vR}
\label{1600}
\eeqn

\beqn
W_3&=&\biggl\la
G_{q_3}\bigl[z_4,z_3;\verqdc,\verqcc;E_3;\vR\bigr]\,
G_{\bar q_2}\bigl[z_4,z_3;\verqdb,\verqcb;E_2;\vR\bigr]\,
G_{q_1}\bigl[z_4,z_3;\verqda,\verqca;E_1;\vR\bigr]
\nonumber\\&\times&
G^*_{q_3}\bigl[z_4,z_3;\vesqda,\vesqcc;\bar E_3;\vR\bigr]\,
G^*_{\bar q_2}\bigl[z_4,z_3;\vesqdb,\vesqcb;\bar E_2;\vR\bigr]\,
G^*_{q_1}\bigl[z_4,z_3;\vesqda,\vesqca;\bar E_1;\vR\bigr]\,
\biggr\ra_{\vR}
\label{1700}
\eeqn
\end{widetext}

They are the solutions of the following equations,
\beqn
&&i\,\frac{\partial W_2}{\partial z_3}=
\biggl[-\frac{\Delta\vec r_3^{\,q_3}}{2 E_3} -
\frac{\Delta\verqcb}{2E_2}-
\frac{\Delta\verqca}{2E_1}+
\frac{\Delta\vec s_3^{\,g}}{2\bar E_g} +
\frac{\Delta\vesqca}{2\bar E_1}
\nonumber\\ &-&
{i\over2}\,\rho_A(b,z_3)\,\Sigma_2(\vec r_3^{\,q_3},\verqcb,\verqca,
\vec s_3^{\,g},\vec s_3^{\,q_1})\biggr]W_2;
\label{1800}
\eeqn

\beqn
&&i\,\frac{\partial W_2}{\partial z_2}=
\biggl[-\frac{\Delta\vec r_2^{\,q_3}}{2 E_3} -
\frac{\Delta\verqbb}{2E_2}-
\frac{\Delta\verqba}{2E_1}+
\frac{\Delta\vec s_2^{\,g}}{2\bar E_g} +
\frac{\Delta\vesqba}{2\bar E_1}
\nonumber\\ &-&
{i\over2}\,\rho_A(b,z_2)\,\Sigma_2(\vec r_2^{\,q_3},\verqbb,\verqba,
\vec s_2^{\,g},\vec s_2^{\,q_1})\biggr]W_2;
\label{1900}
\eeqn

\beqn
&&i\,\frac{\partial W_3}{\partial z_4}=
\biggl[-\frac{\Delta\vec r_3^{\,q_3}}{2 E_3} -
\frac{\Delta\verqcb}{2E_2}-
\frac{\Delta\verqca}{2E_1}+
\frac{\Delta\vec s_3^{\,g}}{2\bar E_g} +
\frac{\Delta\vesqca}{2\bar E_1}
\nonumber\\ &-&
{i\over2}\,\rho_A(b,z_4)\,\Sigma_3(\verqcc,\verqcb,\verqca,
\vesqcc,\vesqcb,\vesqca)\biggr]W_3;
\label{2000}
\eeqn

\beqn
&&i\,\frac{\partial W_3}{\partial z_3}=
\biggl[-\frac{\Delta\vec r_3^{\,q_3}}{2 E_3} -
\frac{\Delta\verqcb}{2E_2}-
\frac{\Delta\verqca}{2E_1}+
\nonumber\\ &+&
\frac{\Delta\vesqcc}{2\bar E_3}+
\frac{\Delta\vesqcb}{2\bar E_2}+
\frac{\Delta\vesqca}{2\bar E_1}
\nonumber\\ &-&
{i\over2}\rho_A(b,z_3)
\Sigma_3(\verqcc,\verqcb,\verqca,
\vesqcc,\vesqcb,\vesqca)\biggr]W_3;
\nonumber\\
\label{2100}
\eeqn

with the boundary conditions,

\beqn
W_2\bigr|_{z_3<z_2} &=& 0
\nonumber\\
W_2\bigr|_{z_3=z_2} &=&
\delta(\verqcc-\verqbc)\,\delta(\verqcb-\verqbb)
\delta(\verqca-\verqba)\,
\nonumber\\ &\times&
\delta(\vesgc-\vesgb)\,
\delta(\vesqca-\vesqba).
\label{2200}
\eeqn

\beqn
&&W_3\bigr|_{z_4<z_3} = 0
\nonumber\\
&&W_3\bigr|_{z_4=z_3} =
\delta(\verqdc-\verqcc)\,\delta(\verqdb-\verqcb)
\delta(\verqda-\verqda)\,
\nonumber\\ &\times&
\delta(\vesqdc-\vesqcc)\,
\delta(\vesqdb-\vesqcb)
\delta(\vesqda-\vesqca).
\label{2300}
\eeqn

The function $\Sigma_2$ is the total cross section of interaction of
the colorless parton ensemble $q_1,\bar q_2, q_3\ \bar q_1,g$ with a
nucleon, where the pairs $q_1\bar q_2$ and $q_1,\bar q_1$ are each
in colorless states, the pair $\bar q_2,q_3$ is a color octet, and
the pair $\bar q_1,g$ is an anti-triplet. Correspondingly,
$\Sigma_3$ is the total cross section for the ensemble $q_1,\bar
q_2,q_3,\ \bar q_1,q_2,\bar q_3$, where each pair $q_1,\bar q_2$,
$\bar q_1,q_2$, and $q_1,\bar q_1$ is colorless, while the pairs
$\bar q_2,q_3$ and $q_2\bar q_3$ are color octets. These cross
sections are derived in \ref{sigmas} and have the form,
\begin{widetext}
\beqn
\Sigma_2 &=&
\sq(\verqb-\verqa)+\sq(\verqa-\vesqa)-
\sq(\verqb-\vesqa) +
{1\over8}\bigl[\sq(\verqc-\verqa) -\sq(\verqc-\verqb)-
\sq(\verqc-\vesqa)\bigr]
\nonumber\\ &+&
{9\over8}\bigl[\sq(\vesqa-\vesg)-
\sq(\verqa-\vesg) +
\sq(\verqb-\vesg) +
\sq(\verqc-\vesg)\bigr];
\label{2400}
\eeqn

\beqn
\Sigma_3 &=&
\sq(\verqa-\verqb)+\sq(\vesqa-\vesqb)-
\sq(\verqa-\vesqb) - \sq(\vesqa-\verqb)+
\sq(\verqa-\vesqa)+\sq(\verqb-\vesqb)
+\nonumber\\ &+&
\sq(\verqc-\vesqc)+
{1\over8}\bigl[\sq(\verqc-\verqa) +\sq(\vesqa-\vesqc)+
\sq(\verqb-\vesqc)+\sq(\vesqb-\verqc) -
\sq(\verqc-\vesqa)
\nonumber\\&-&
 \sq(\verqa-\vesqc) -
\sq(\verqb-\verqc)-\sq(\vesqb-\vesqc)
\bigr] .
\label{2500}
\eeqn

\end{widetext}

The equations~(\ref{800})-(\ref{850}) and (\ref{1800})-(\ref{2100})
have been solved in \ref{W}, in the approximation of
Eq.~(\ref{1100}) and for a constant nuclear density.

\section{The three parts of the cross section}\label{3parts}

In order to discriminate between production of the pion within or
outside the nucleus, we rely on the approximation of constant
nuclear density, $\rho_A(b,z)=\rho_0\Theta(L^2-z^2)$, where
$L=\sqrt{R_A^2-b^2}$. Then we split the amplitude
equation~(\ref{100}) in two parts, $M=M_1+M_2$, corresponding to
pion production outside, or inside the nucleus, i.e. $z_2$
-integration within intervals $L<z_2<z_3$ and $z_1<z_2<L$
respectively. Correspondingly, the cross section
equation~(\ref{1300}) contains three terms, \beq
\sigma=\sigma_1+\sigma_2+\sigma_3, \label{2550} \eeq which are the
amplitudes $M_1$, $M_2$ squared, and their interference,
respectively. These terms  correspond to the following splitting of
the integrations over $z_2$ and $z_3$ in (\ref{1300}), \beq
\int\limits_{z_1}^{z_4}dz_2\int\limits_{z_2}^{z_4}dz_3=
\int\limits_{L}^{z_4}dz_2\int\limits_{z_2}^{z_4}dz_3+
\int\limits_{z_1}^{L}dz_2\int\limits_{z_2}^{L}dz_3+
\int\limits_{z_1}^{L}dz_2\int\limits_{L}^{z_4}dz_3. 
\label{2600}
\eeq

In what follows we consider cross sections integrated over
transverse momenta of the pion and recoil quark,
\beq
\frac{d\sigma_i}{dz_h}=\int d^2p_\pi\,d^2p_3\,
\frac{d\sigma_i}{d^2p_\pi\,d^2p_3\,dz_h}, 
\label{2700}
\eeq
where $i=1,\ 2,\ 3$.

Later, the results of numerical calculations will show that the interference term is negative, $\sigma_3<0$. This can be understood on a much simplified example of an "empty" nucleus, i.e. free propagation of particles. In this case the amplitude of the fragmentation process $q_1\to q_1\bar q_2q_3$ is proportional to the value,
\beqn
\mathcal{M}&=&\lim\limits_{z_+\to\infty}
\int\limits_{z_1}^{z_+} dz\,
\exp\Bigl[-i(\Delta-io)z\Bigr]
\nonumber\\ &\times&
G_g(z,z_1;\vec p_g)
G_{q_1}(z_+,z;\vec p_1)
G_{\bar q_2}(z_+,z;\vec p_2),
\label{2710}
\eeqn
where $\vec p_g=\vec p_1+\vec p_2$;
\beqn
\Delta &=& \frac{m_q^2}{2E_1}+
\frac{m_q^2}{2E_2};
\nonumber\\
G_g(z,z_1;\vec p_g) &=&
\exp\left[-\frac{i\vec p_g^{\,2}(z-z_1)}{2E_g}\right];
\nonumber\\
G_{q_1}(z_+,z;\vec p_1) &=&
\exp\left[-\frac{i\vec p_1^{\,2}(z_+-z)}{2E_1}\right];
\nonumber\\
G_{\bar q_2}(z_+,z;\vec p_2) &=&
\exp\left[-\frac{i\vec p_2^{\,2}(z_+-z)}{2E_2}\right];
\label{2720}
\eeqn

This amplitude can be represented as,
\beq
\mathcal{M}=\lim\limits_{z_+\to\infty}
e^{i\varphi}
\int\limits_{z_1}^{z_+} dz\,
e^{-i\vartheta z},
\label{2730}
\eeq
where
\beqn
\vartheta &=&
\frac{m_q^2+p_1^2}{2E_1}+
\frac{m_q^2+p_2^2}{2E_2}-
\frac{(\vec p_1+\vec p_2)^2}{2E_g}
-io; \label{2740}\\
\varphi &=&
\left(\frac{p_1^2}{2E_1}+
\frac{p_2^2}{2E_2}\right)z_+
- \frac{(\vec p_1+\vec p_2)^2}{2E_g}\,z_1.
\label{2745}
\eeqn

Now we can split the amplitude into two terms, $\mathcal{M}=\mathcal{M}_{in}+\mathcal{M}_{out}$, corresponding to gluon decay inside ($z_1<z<\bar z$) and outside ($\bar z<z<z_+$) the nucleus respectively. Then from (\ref{2730}) we get,
\beqn
\mathcal{M}_{in} &=&
\frac{e^{i\varphi}}{i\vartheta}\left(e^{-i\vartheta\bar z} -
e^{-i\vartheta z_1}\right);
\label{2750}\\ 
\mathcal{M}_{out} &=&
\frac{e^{i\varphi}}{i\vartheta}\left(e^{-i\vartheta z_+} -
e^{-i\vartheta \bar z}\right).
\label{2760}
\eeqn

At $z_+\to\infty$ the first term in $\mathcal{M}_{out}$ vanishes,
$e^{-i\vartheta z_+}\to 0$, because of the imaginary term in $\vartheta$, Eq.~(\ref{2740}).
So we get,
\beqn
\left|\mathcal{M}_{out}\right|^2 &=&
\frac{1}{\vartheta^2};
\label{2770}\\
\left|\mathcal{M}_{in}\right|^2 &=&
\frac{2}{\vartheta^2}
\Bigl\{1-\cos\bigl[\vartheta(\bar z-z_1)\bigr]\Bigr\};
\label{2780}\\
2\Re \left(\mathcal{M}_{in}
\mathcal{M}^*_{out}\right) &=&
- \frac{2}{\vartheta^2}
\Bigl\{1-\cos\bigl[\vartheta(\bar z-z_1)\bigr]\Bigr\}
\nonumber\\ &=& 
- \left|\mathcal{M}_{in}\right|^2.
\label{2790}
\eeqn
Thus, we conclude that the interference term (\ref{2790}) is negative and exactly cancels the inside production term. The cross section in this case is given solely by the outside production.

Of course, these simple results are valid only for hadronization in vacuum ("empty" nucleus). Presence of a medium breaks down these simple relations and makes the calculation of different terms in the cross section Eq.~(\ref{2550}), performed below, much more complicated. Nevertheless, the negative sign of the interference term will be preserved.

\subsection{Pion production outside the nucleus}

We start with the first term $\sigma_1$, which dominates at high
energy $E$, and is the easiest one to calculate, since in
this case the functions $W_{1,2}$ contain just products of Green
functions for free propagation of quarks in vacuum. So the
integration over longitudinal coordinates in (\ref{1300}) and
transverse momenta in (\ref{2700}) can be performed analytically.

\begin{widetext}
\beqn
\frac{d\sigma_1}{dz_h d^2b}&=&
A\,\rho_0 E^2\,F_2^N(x,Q^2)\,
\frac{z^2(1-z)^2}{x}
\int\limits_{-L}^L dz_1
\int d^2r_1^g d^2s_1^g\,
\tilde\Gamma(\verga)\,\tilde\Gamma(\vesga)
\int d^2r_2^g d^2r_2^{q_1} d^2s_2^g d^2s_2^{q_1}
d\alpha\,d\bar\alpha
\nonumber\\ &\times&
W_1(L,z_1;\vergb,\verqba,\vesgb,\vesqba;
\verga,\verga,\vesga,\vesga)
\Phi(\vergb,\verqba;\vesgb,\vesqba),
\label{2800}
\eeqn
with the new notation,
\beqn
\Phi(\vergb,\verqba;\vesgb,\vesqba)&=&
\delta\bigl(\vec R_2-\vec S_2\bigr)
\int d^2\rho\,d^2\tau\,
K_0(m_q\rho)\,K_0(m_q\tau)\,
\Phi_\pi\left(\vec r_2+\frac{1-z_h}{1-\alpha z_h}\,\vec\rho\right)\,
\Phi_\pi^*\left(\vec s_2+\frac{1-z_h}{1-\bar\alpha z_h}\,\vec\tau\right)
\nonumber\\ &\times&
\delta\left(\alpha\vec r_2-\frac{1-\alpha}{1-\alpha z_h}\,
\vec\rho-\bar\alpha\vec S_2+
\frac{1-\bar\alpha}{1-\bar\alpha z_h}\,\vec\tau\right),
\label{2900}
\eeqn
\end{widetext}
where $\vec r_i=\vec r_i^{\,g}-\vec r_i^{\,q_1}$; $\vec s_i=\vec s_i^{\,g}-\vec s_i^{\,q_1}$, and
\beqn
\vec R_i&=&\frac{E_g\,\vec r_i^{\,g}+E_1\,\vec r_i^{\,q_1}}
{E_g+E_1};
\nonumber\\
\vec S_i&=&\frac{\bar E_g\,\vec s_i^{\,g}+
\bar E_1\,\vec s_i^{\,q_1}}
{\bar E_g+\bar E_1},
\label{3000}
\eeqn
($i=1,\ 2$) are  the intrinsic separations in the $q_1-g$ pairs,
and the coordinates of their centers of gravity, respectively;
$E_g+E_1=\bar E_g+\bar E_1=E$.

We also introduce the following combinations,
\beqn
\vec R_i^{\,+}&=&{1\over2}\,\bigl(\vec R_i+\vec S_i\bigr),
\nonumber\\
\vec R_i^{\,-}&=&\vec R_i-\vec S_i.
\label{3100}
\eeqn
The Jacobian for transition to the new coordinates is one, so,
\beq
d^2r_2^g\,d^2r_2^{q_1}\,
d^2r_2^g\,d^2r_2^{q_1}=
d^2R_2^+\,d^2R_2^-\,
d^2r_2\,d^2s_2.
\label{3200}
\eeq

It turns out that the function $W_1$ factorizes in the new coordinates (see \ref{W}),
\beqn
&&W_1\left(L,z_1;\vec r_2^{\,g},\vec r_2^{\,q_1};
 \vec s_2^{\,g},\vec s_2^{\,q_1};
 \vec r_1^{\,g},\vec r_1^{\,q_1};
 \vec s_1^{\,g},\vec s_1^{\,q_1}\right)
  \nonumber\\ &=&
 \left[\frac{E}{2\pi(L-z_1)}\right]^2
F\left(L,z_1;\vec R_2^{\,-},\vec r_2,\vec s_2;
\vec R_1^{\,-},\vec r_1,\vec s_1\right).
\nonumber\\ &\times&
 \exp\left[\frac{iE}{L-z_1}\,
 \left(\vec R_2^{\,+}-\vec R_1^{\,+}\right)
\left( \vec R_2^{\,-}-
 \vec R_1^{\,-}\right)\right]
  \label{3300}
\eeqn
Taking also into account that in (\ref{2800}) $\verga=\verqaa$ and
$\vesga=\vesqaa$ (i.e. $\vec r_1=\vec s_1=0$), we arrive at the
relation,
\beqn
&&\int d^2r_1^g\,d^2s_1^g\, d^2R_2^+\,d^2R_2^-\,
\tilde\Gamma(\verga)\,\tilde\Gamma(\vesga) \nonumber\\ &\times&
W_1(L,z_1;\vergb,\verqba,\vesgb,\vesqba;
\verga,\verga,\vesga,\vesga)\, \delta\left(\vec R_2-\vec S_2\right)
\nonumber\\ &=& \int d^2R_1^+\, \left|\tilde\Gamma\left(\vec
R_1^{\,+}\right) \right|^2\, \tilde F(L,z_1;\vec r_2,\vec s_2),
\label{3400} \eeqn where \beqn &&\tilde F(L,z_1;\vec r_2,\vec s_2)
\nonumber\\ &=& F\left(L,z_1;\vec R_2^{\,-},\vec r_2,\vec s_2; \vec
R_1^{\,-},\vec r_1,\vec s_1\right)_{ \scriptsize
\begin{array}{c}
\vec R_1^{\,-}=\vec R_2^{\,-}=0  \\
\vec r_1=\vec s_1=0
   \end{array}}
\label{3500}
\eeqn

Thus the cross section equation~(\ref{2800}) gets the form,
\beqn
\frac{d\sigma_1}{dz_h d^2b} &=&
A\,N z_h^2(1-z_h)^2 \rho_0
\int\limits_{-L}^Ldz_1
\int d^2r_2 d^2s_2 d\alpha d\bar\alpha
\nonumber\\ &\times&
\tilde F(L,z_1;\vec r_2,\vec s_2;\alpha,\bar\alpha)\,
\tilde\Phi(\vec r_2,\vec s_2),
\label{3600}
\eeqn
where
\beqn
&&\tilde\Phi(\vec r_2,\vec s_2)=
\int d^2\rho d^2\tau\,K_0(m_q\rho)\,K_0(m_q\tau)
\nonumber\\ &\times&
\Phi_\pi\left(\alpha;\vec r_2+\frac{1-z_h}{1-\alpha z_h}\,\vec\rho\right)\,
\Phi_\pi\left(\bar\alpha;\vec s_2+
\frac{1-z_h}{1-\bar\alpha z_h}\,\vec\tau\right)
\nonumber\\&\times&
\delta\left(\alpha\vec r_2-\frac{1-\alpha}{1-\alpha z_h}\,\vec\rho -
\bar\alpha\vec s_2+
\frac{1-\bar\alpha}{1-\bar\alpha z_h}\,\vec\tau\right),
\label{3700}
\eeqn
and
\beq
N=\frac{F_2(x)}{x}\, E^2\int d^2r
\left|\tilde\Gamma(\vec r)\right|^2.
\label{3800}
\eeq

In order to simplify the calculations we assume a factorized form of
the pion light-cone wave function, $\Phi_\pi(\alpha,\vec
r)=\varphi(\alpha)\,\phi(\vec r)$, and a Gaussian dependence on
quark separation,
\beq
\phi(r)\propto\exp\left(-\frac{\xi}{2}\,r^2\right), 
\label{3850}
\eeq
where $\xi$ is related to the mean pion charge radius squared,
$\xi=3/8\la r_{ch}^2\ra$.

Perturbative fragmentation of quarks to pions in $e^+e^-$
annihilation and DIS was calculated by Berger \cite{berger} in the
limit of $(1-z_h)\ll1$. The pion wave function was maximally simplified
assuming that $\varphi(\alpha)=\delta(\alpha-1/2)$ and fixing at
zero the relative $\bar qq$ momentum. This simplifies the
calculations considerably, since the function $\tilde F$ can be
obtained analytically,
\beqn &&\tilde F(L,z_1;\vec r_2,\vec
s_2;\alpha,\bar\alpha)_{\alpha=\bar\alpha} \nonumber\\ &=&
\left(\frac{\varepsilon}{2\pi(L-z_1)}\right)^2
\exp\biggl[\frac{i\,\varepsilon}{2(L-z_1)}\,(\vec r_2^{\,2}- \vec
s_2^{\,2}) \nonumber\\ &-& {1\over6}\,\rho_0\,C(z_h,\alpha)(\vec
r_2-\vec s_2)^2(L-z_1)\biggr], \label{3900}
\eeqn
where $ \varepsilon=\alpha z_h(1-\alpha z_h)E$;
\beq C(z_h,\alpha)=C\left(1+\alpha^2z_h^2+
\frac{\alpha z_h}{4}\right).
\label{4000}
\eeq

Notice that in this case the expressions for $W_2$ and $W_3$,
equations~(\ref{1600}) and (\ref{1700}), also are much simplified.

We can perform the integration over the transverse coordinates and
momenta using the integral representation for the modified Bessel
functions,
\beqn &&K_0(m_q\rho)\,K_0(m_q\tau)=
{1\over2}\int\limits_{-1}^1\frac{dv}{1-v^2} \int\limits_0^\infty dw
\nonumber\\ &\times& \exp\left[-\frac{m_q^2}{2w}
\left(\frac{\rho^2}{1+v}+\frac{\tau^2}{1-v}\right) -w\right]
\label{4100}
\eeqn
Then in the case of equal sharing of longitudinal
momentum by the pion quarks we arrive at a simple result,
\beqn
\frac{d\sigma_1}{dz_h d^2b}&=&
A\,N\,z_h^2(1-z_h)^2\rho_0\int\limits_{-L}^L dz_1
\int\limits_0^\infty dw \nonumber\\ &\times& \int\limits_{-1}^1 dv\,
\frac{e^{-w}}{m_q^2 +aD(w,v,z_1)}, \label{4200} \eeqn where \beqn
D(w,v,z_1) &=& (1-v^2)w\left(\frac{1-z_h}{1-x}\right)^2 \nonumber\\
&+& (1-x)^2m_q^2\left(ua+{m_q^2\over w}\right)
\left(\frac{L-z_1}{\varepsilon}\right)^2 \nonumber\\ &+&
{2\over3}\,C\rho_0\left(1+x^2+{x\over4}\right)
(L-z_1)^3\left(\frac{m_q}{\varepsilon}\right)^2 \nonumber\\ &+&
2i\,v(1-z_h)\, \frac{m_q^2(L-z_1)}{\varepsilon}. 
\label{4300}
\eeqn

Notice that although the function $D(w,v,z_1)$ is complex, the
expression (\ref{4200}) is real.

In the limit of $\xi\to0$ in Eq.~(\ref{3850}) the cross section
Eq.~(\ref{4200}) does not depend any more on the interaction with
the medium, which is characterized by the constant $C$. Thus in the
Berger model for fragmentation ($\alpha=1/2;\
\la\kappa^2\ra=\xi=0$), the interaction of the quark and gluon with
the medium does not affect the value of the cross section
$\sigma_1$,Eq.~(\ref{4200}), and only modifies the transverse
momentum distribution, which is an effect beyond the scope of this
study.

In another limiting case $\Phi_\pi(\alpha,\vec\kappa)=\varphi(\alpha)\,
\delta(\vec\kappa)$ the cross section gets the form,
\beqn
\frac{d\sigma_1}{dz_h d^2b}&=&
A\,N\,z_h^2(1-z_h)^2\rho_0\int\limits_{-L}^L dz_1
\int\limits_0^\infty dw
\int\limits_{0}^1 d\alpha
\int\limits_{0}^1 d\bar\alpha
\nonumber\\ &\times&
\frac{(1-\alpha z_h)(1-\bar\alpha z_h)\varphi(\alpha)\varphi(\bar\alpha)}
{D_1\,D_2\,D_3}\,
\gimel\,\aleph\,e^{-w}.
\label{4400}
\eeqn
 We use here the following notation,
 \beqn
 D_1&=& (1-\alpha)^2t\,w+m_q^2(1-\alpha z_h)^2u;\nonumber\\
 D_2&=& (1-\bar\alpha)^2t\,w+m_q^2(1-\bar\alpha z_h)^2u;\nonumber\\
 D_3&=& \cos\left[\omega_1(L-z_1)\right]\,
\cos\left[\omega_2(L-z_1)\right];
\label{4500}
\eeqn
\beq
\gimel=\varepsilon\bar \varepsilon\omega_1\omega_2\,
\cot\left[\omega_1(L-z_1)\right]\,
\cot\left[\omega_2(L-z_1)\right];
\label{4600}
\eeq
 \beqn
 \aleph&=& \frac{i}{2(1-\mu_1\mu_2)}\,
\left\{\bar \varepsilon\omega_2\beta^2\cot\left[\omega_1(L-z_1)\right]
\right.\nonumber\\ &-& \left.
\varepsilon\omega_1\gamma^2\cot\left[\omega_1(L-z_1)\right]
\right\},
\label{4700}
\eeqn
where $\beta=\alpha-\nu\bar\alpha$; $\gamma=\bar\alpha-\mu\alpha$;
$\varepsilon=E\alpha z_h(1-\alpha z_h)$;
$\bar\varepsilon=E\bar\alpha z_h(1-\bar\alpha z_h)$;
$\omega_1=\sqrt{-i\lambda_1}$;
$\omega_2=\sqrt{-i\lambda_2}$;
\beqn
\lambda_{1,2}&=&C\,\rho_0\,
\frac{\sqrt{(\bar\varepsilon a+\varepsilon b)^2-4\varepsilon\bar\varepsilon c^2}\pm
\bar\varepsilon a\mp\varepsilon b}{2\varepsilon\bar\varepsilon};
\nonumber\\
\bar\varepsilon\mu &=& \varepsilon\nu =
\frac{\bar\varepsilon a+\varepsilon b
-\sqrt{(\bar\varepsilon a+\varepsilon b)^2-4\varepsilon\bar\varepsilon c^2}}{2\,c};
\label{4800}
\eeqn
 \beqn
 a&=&(1-\alpha z_h)^2+{9\over4}\,\alpha z_h;
 \nonumber\\
 b&=&(1-\bar\alpha z_h)^2+{9\over4}\,\bar\alpha z_h;
 \nonumber\\
 c&=&(1-\alpha z_h)(1-\bar\alpha z_h)+
 {9\over8}\,(\alpha z_h+\bar\alpha z_h).
 \label{4900}
 \eeqn

\begin{figure}[htb]
 \centerline{\includegraphics[width=9cm]{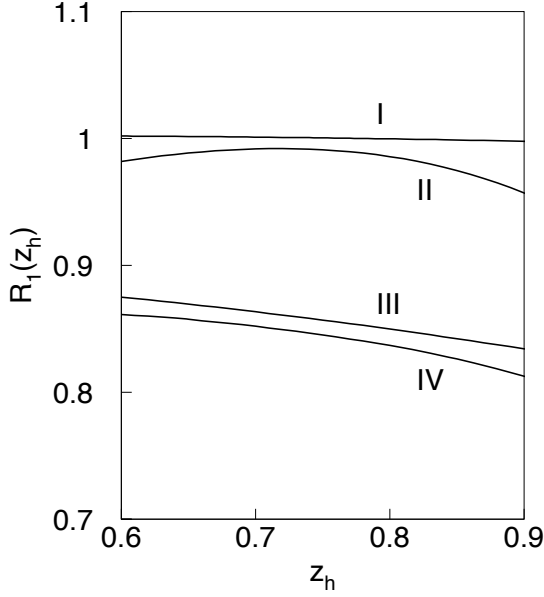}}
 \caption{\label{comparison}
 Comparison of ratios $R_1(z_h)$, Eq.~(\ref{5000}), for pre-hadron
production outside the nucleus, for different models for the pion
light-cone wave function. The four variants I-IV (see text) differ
by absence or presence of longitudinal and transverse motion of the
valence quarks in the pion. The calculations are done for lead at
$E=10\GeV$.
Since the recoil quark should be ultra-relativistic in order
to rely on the Green function method, we
restricted the range of $z_h < 0.9$.}
 \end{figure}

 Nuclear effects for this part of the cross section, $\sigma_1$, are
 shown in Fig.~\ref{comparison} in the form of ratio,
 \beq
 R_1(z_h)=\frac{d\sigma_1/dz_h}
{d\sigma_1(C=0)/dz_h}, 
\label{5000} 
\eeq
where both the numerator and denominator are the cross sections on the nucleus integrated over impact parameter, however in the denominator we eliminate the influence of the medium fixing the imaginary part of the light-cone potential $C=0$, so the quark and gluon propagate like in vacuum. The nuclear cross section, here in the numerator and in what follows, is calculated with $C=3$. This value agrees with extrapolation of the saturated cross section \cite{kst2} down to medium high energies, as well as agrees with data on nuclear broadening of transverse momentum \cite{kns07}.

We performed calculations for four cases:\\
I. $\Phi_\pi(\alpha,\vec\kappa)\propto\delta(\alpha-1/2)\,\delta(\vec\kappa)$ (Berger
approximation);\\
II. $\Phi_\pi(\alpha,\vec\kappa)=\varphi(\alpha)\,\delta(\vec\kappa)$, where
$\varphi(\alpha)\propto\alpha(1-\alpha)$;\\
III. $\Phi_\pi(\alpha,\vec\kappa)=\delta(\alpha-1/2)\,\phi(\vec\kappa)$, where
$\phi(\vec\kappa)\propto\exp(-\xi\kappa^2/2)$;\\
IV. $\Phi_\pi(\alpha,\vec\kappa)=\varphi(\alpha)\,\phi(\vec\kappa)$, where
$\varphi(\alpha)\propto\alpha(1-\alpha)$ \cite{radyushkin,stan}, $\phi(\vec\kappa)\propto\exp(-\xi\kappa^2/2)$.

Comparing curves I, II with III, IV in Fig.~\ref{comparison} one can
conclude that the transverse motion of quarks
($\xi=\la\kappa^2\ra\neq0$) significantly affects the
nucleus-to-proton ratio. At the same time,  the
$\alpha$-distribution, i.e. longitudinal motion of quarks in the
pion, has almost no influence on the nuclear effects. Indeed, the
curves III and IV are nearly very close to each other. In what follows
we assume that $\sigma_{2,3}$ are also insensitive to the form of
the $\alpha$-distribution, so we will continue our calculations in
the approximation III,
$\Phi_\pi(\alpha,\vec\kappa)=\delta(\alpha-1/2)
\exp(-\xi\kappa^2/2)$.

 \subsection{Pion production inside the nucleus}

 The second term in the cross section Eq.~(\ref{5500}), which corresponds to gluon decay
 inside the nucleus,
 has the form,
 \beqn
 &&\frac{d\sigma_2}{dz_h d^2b} =
 \pi\,A\,N\,\rho_0\, \Re
 \int\limits_{-L}^L dz_1
 \int\limits_{z_1}^L dz_2
  \int\limits_{z_2}^L dz_3
  \nonumber\\ &\times&
  \frac{-i\mathcal{E}\omega
  \exp\left[-\frac{im_q^2}{2\mathcal{E}}\,(z_3-z_2)\right]}
  {E^2\left(1+\frac{\xi}{\varepsilon^{\,2}}\,H_2\right)
  \sin\large[\omega(z_3-z_2)\large]}.
 \label{5100}
 \eeqn
 Here
 \beqn
 H_2&=&{2\over3}\,\rho_0
 \large[(c_1-c_2)(z_2-z_1)^3+
 (c_3-c_2)(z_3-z_1)^3
 \nonumber\\ &+&
 c_3(L-z_1)^3\large]+
 2i\,\mathcal{E}\lambda(2x+\lambda)(z_3-z_2)
 \nonumber\\ &-&
 2i\,\mathcal{E}\omega
 \Bigl\{\bigl[(x+\lambda)^2(z_3-z_1)^2+\lambda^2(z_2-z_1)^2\bigr]
 \nonumber\\ &\times&
 \cot\Large[\omega(z_3-z_2)\Large]
 \nonumber\\ &-&
  \frac{2(x+\lambda)\lambda(z_3-z_1)(z_2-z_1)}
{\sin\Large[\omega(z_3-z_2)\Large]}
 \nonumber\\ &+&
 4\xi(1-x)^2(L-z_1)^2,
 \label{5200}
 \eeqn
 and
 \beqn
 \mathcal{E}&=&(1-z_h)\frac{x\,E}{1-x};\nonumber\\
  x&=&z_h/2;\nonumber\\
 \lambda&=&{q\over p}; \nonumber\\
 p &=& \frac{4-17x+22x^2}{4(1-x)^2};\nonumber\\
 q &=& \frac{4-8x-5x^2}{4(1-x)};\nonumber\\
 c_1 &=& \left(1+{x\over4}+x^2\right)\,C;\nonumber\\
 c_2 &=&  \left(1+{x\over4}+x^2-{q^2\over p}\right)\,C;\nonumber\\
 c_3&=&4(1-x)^2C.
 \label{5300}
  \eeqn
 One can see that even in the limit $\xi\to0$ the cross section Eq.~(\ref{5100})
 is still sensitive to the constant $C$ due to the presence of $\sin\Large[\omega(z_3-z_2)\Large]$ in (\ref{5200}).

\begin{figure}[htb]
 \centerline{\includegraphics[width=7cm]{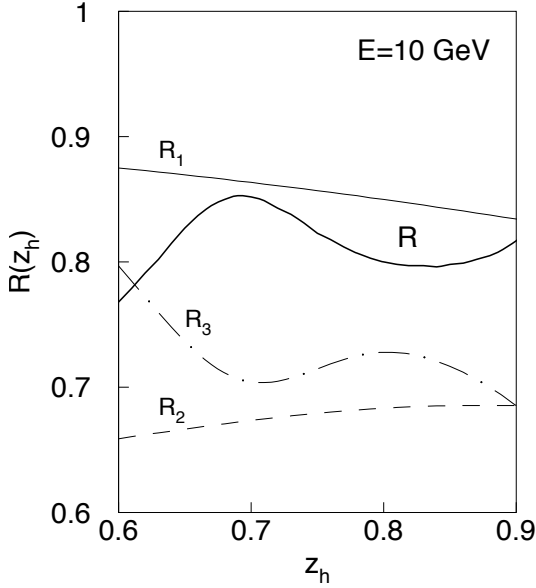}}
 \caption{\label{1-2-3} Lead-to-proton target ratios for the cross sections corresponding to
 pre-hadron production outside ($R_1$) or inside ($R_2$) the nucleus in both
 amplitudes.
 The ratio for inside-outside interference is shown by the curve indicated by $R_3$.
 The ratio of the full cross sections $\sigma_1+\sigma_2+\sigma_3$, Eq.~(\ref{5900}),
 is depicted by the solid curve indicated by $R$. Notice that the interference term $\sigma_3$ is negative.}
 \end{figure}

The result for the ratio
 \beq
 R_2(z_h)=\frac{d\sigma_2/dz_h}
 {d\sigma_2(C=0)/dz_h},
 \label{5400}
 \eeq
 calculated for lead at $10\GeV$ is depicted in Fig.~\ref{1-2-3}.
As one could expect, this contribution is more suppressed, since the gluon decays inside the nucleus producing the colorless pre-hadron which propagates and attenuates in the medium.

 \subsection{Interference of amplitudes \boldmath$M_1$ and $M_2$}

Eventually, the third term in Eq.~(\ref{2550}), corresponding to
interference of the two amplitudes, after integration over
transverse variables gets the form,
 \beqn
&& \frac{d\sigma_3}{dz_h d^2b}=
 \pi\,A\,N\,\rho_0\,\Re
 \int\limits_{-L}^L dz_1
 \int\limits_{z_1}^L dz_2
 \int\limits_0^\infty \frac{dw}{w}\,e^{-w}
 \nonumber\\ &\times&
 \frac{-i\mathcal{E}\omega
  \exp\left[-\frac{im_q^2}{2\mathcal{E}}\,(L-z_2)\right]}
  {\left[\left(1+\frac{\xi}{\varepsilon^{2}}\,H_3\right)A_3-\xi\,B_3^2\right]
\sin\large[\omega(L-z_2)\large]}.
 \label{5500}
 \eeqn
 Here
 \beqn
 H_3&=& {m_q^2\over w}\,(1-x)^2(L-z_1)^2
 \nonumber\\ &+&
 {2\over3}\,\rho_0\Large[(c_1-c_2)(L-z_2)^3
 +c_2(L-z_1)^3\Large]
 \nonumber\\ &+&
 4\xi(1-x)^2(L-z_1)^2 +2i\mathcal{E}\lambda^2(L-z_2)
 \nonumber\\ &-&
 2i\mathcal{E}\omega\biggl\{\bigl[(L-z_1)^2+(z_2-z_1)^2\bigr]
 \cot\large[\omega(L-z_2)\large]
 \nonumber\\ &-&
 \frac{(L-z_1)(z_2-z_1)}{\sin\large[\omega(L-z_2)\large]}\biggr\} +
 2i\varepsilon(1-z_h)(L-z_1);
 \label{5600}
 \eeqn
 \beqn
 A_3={m_q^2\over w}-2i\mathcal{E}\omega
 \cot\Large[\omega(L-z_2)\Large];
 \label{5700}
 \eeqn
 \beqn
 B_3 &=& \frac{m_q^2}{w\varepsilon}(1-x)(L-z_1) +
 2i\left(\frac{1-z_h}{1-x}\right)
 \nonumber\\ &-&
 2i\mathcal{E}\omega\lambda\biggl\{(L-z_1)
 \cot\Large[\omega(L-z_2)\Large]
 \nonumber\\ &-&
 \frac{z_2-z_1}{\sin\Large[\omega(L-z_2)\Large]}\biggr\}.
 \label{5800}
 \eeqn

 The values of $R_3(z_h)$, defined similarly to Eqs.~(\ref{5000}),
 (\ref{5400}), are depicted in Fig.~\ref{1-2-3}.
  The solid curve in this figure presents the final results for the ratio of all terms in (\ref{5500}),
 \beq
 R(z_h)=\frac{\left(\frac{d\sigma_1}{dz_h}+
 \frac{d\sigma_2}{dz_h}+\frac{d\sigma_3}{dz_h}\right)}
 {\left(\frac{d\sigma_1}{dz_h}+
 \frac{d\sigma_2}{dz_h}+\frac{d\sigma_3}{dz_h}\right)_{C=0}}
 \label{5900}
 \eeq

\begin{figure}[htb]
 \centerline{\includegraphics[width=9cm]{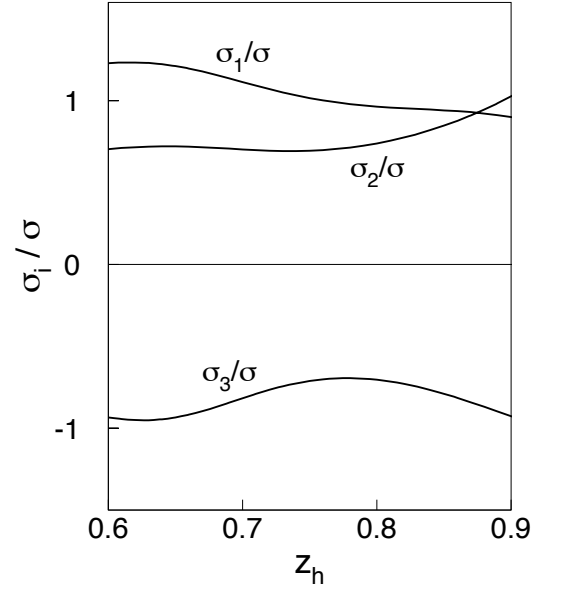}}
 \caption{\label{3sigmas} Fractions $\sigma_i/(\sigma_1+\sigma_2+\sigma_3)$ calculated for lead at $E= 10\GeV$. }
 \end{figure}

Fig.~\ref{comparison} does not contain information about relative contribution of different terms in (\ref{5900}) to the cross section. To show that we depicted the fractions $\sigma_i/(\sigma_1+\sigma_2+\sigma_3)$ in Fig.~\ref{3sigmas}.
 All three terms in the numerator of (\ref{5900}) are of the same order, but the last one presenting interference, is negative. The latter was expected according to the calculation performed in Sect.~\ref{3parts} for hadronization in vacuum. We see that at $10\GeV$ the fractions of the cross section corresponding to production inside and outside the nucleus are about equal, while the former is more suppressed according to Fig.~\ref{1-2-3}.  This is, however, a classical interpretation, the inside-outside interference term $\sigma_3$ does not allow to classify events this way.

\begin{figure}[htb]
 \centerline{\includegraphics[width=9cm]{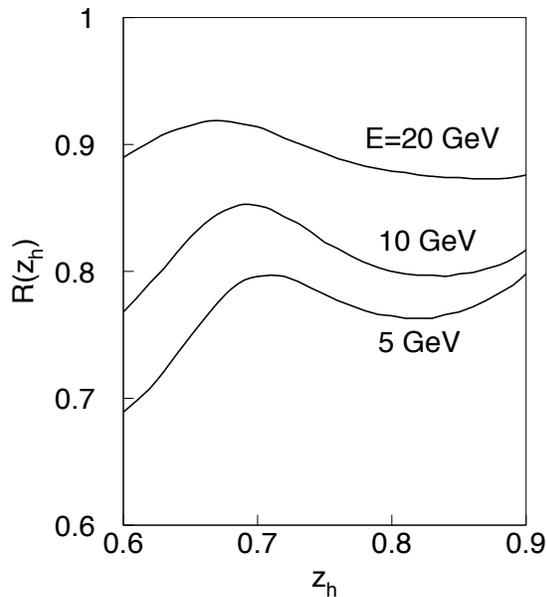}}
 \caption{\label{e-dep} Ratio $R(z_h)$ defined in (\ref{5900}) for lead at different photon
 energies $E= 5,\ 10,\ 15\GeV$.}
 \end{figure}

 The nuclear effects represented by the ratio $R(z_h)$ depend on the photon
 energy, and the higher the energy is, the weaker is the nuclear suppression.
 This is the obvious manifestation of color transparency \cite{zkl}: the initially 
 small quark-gluon separation (see Sect.~\ref{q2-dep})
 is evolving slower at high energy due to Lorentz time dilation. 
The energy dependence is illustrated in Fig.~\ref{e-dep} by some examples.

\section{Main results and observations}\label{results}

This paper presents the first attempt to describe hadronization of a
parton propagating through a medium on a fully quantum-mechanical
basis. For hadronization in vacuum we employ the Berger model
\cite{berger} of perturbative fragmentation, improved in
\cite{born}. This mechanism, imbedded in a nuclear environment, is
illustrated in Fig.~\ref{gamma-a}, and the associated space-time
development in Fig.~\ref{1-2-3}. We employed the path-integral
formulation \cite{feynman} of quantum mechanics, which describes
propagation of partons and partonic ensembles in terms of the
light-cone Green function formalism. This technique properly
includes all phases and takes care of all coherence phenomena,
including formation of hadronic wave functions and color
transparency.

The important observations of the paper can be summarized as follows.
\begin{itemize}

\item
Contrary to the usual expectation based on classical intuition, even
if the radiated gluon always decays outside the medium and the
produced $\bar q_2 q_3$ pre-hadron has no final state interactions,
there is a considerable nuclear suppression for pion production, as
is demonstrated by the ratio $R_1(z_h)$ in Fig.~\ref{1-2-3}. Notice
that no energy loss effect or final state absorption are involved in
this result. The suppression is caused by multiple interactions of
the partons in the medium affecting the overlap of the pre-hadron
and pion wave functions, even if the pre-hadron is produced far away
from the nucleus. The effect of nuclear suppression is subject to color transparency and is controlled by the size of the effective dipoles. The latter is evolving starting from a very small separation $\sim1/Q$ in the hard reaction initiating the jet. The magnitude and energy dependence of nuclear suppression is similar to what is known for electroproduction of $\rho$-mesons in the regime of short coherence length \cite{ct1,ct2}. In that case a dipole is also produced in a small-size configuration and then is evolving with a speed dependent on energy.   

\item
Although much more involved, the effect of nuclear suppression 
of pre-hadrons produced outside the nucleus is in some respect analogous to gluon shadowing (there are also important differences). 
Indeed, in the nuclear rest frame gluon shadowing looks like suppression 
of gluon radiation by multiple interactions 
\cite{kst1,kst2} (Landau-Pomeranchuk effect). 
In this case there are no colorless objects to be 
absorbed in the nucleus, yet the production rate of gluons is affected by 
the medium. In this case gluons radiated inside and outside the nucleus
also interfere.

\item
The novel feature related to the quantum-mechanical treatment of the
problem, is the production of the pre-hadron both inside and outside
the nucleus. This is analogous to the Twin Slit Interference
Experiment in quantum mechanics when a particle propagates
simultaneously through both slits. Interference of the amplitudes
with inside/outside pre-hadron production has a considerable effect
on the nuclear absorption. This interference term in the cross section is large and negative, as is explained in Sect.~\ref{3parts} on the example of hadronization in vacuum.
It is not a surprise that the possibility
of pre-hadron production inside the nucleus leads to  more
suppression due to attenuation of the colorless pre-hadron $\bar q_2
q_1$.

\item
Suppression of hadrons should be much stronger in the case of a
dense medium created in heavy ion collisions. This effect is
completely missed in calculations based on the energy loss scenario
\cite{e-loss}. In fact, it should account for a substantial part of
high-$p_T$ hadron suppression observed in heavy ion collisions
\cite{quenching}. This may also explain why the observed
suppression, when is related solely to energy loss, demands an
unrealistically high density of gluons radiated in heavy ion
collisions \cite{high-density}.

\end{itemize}

While the performed analysis highlights the novel features of
in-medium fragmentation brought by a rigorous quantum-mechanical
treatment of the process, it is still not sufficiently realistic to
be compared with data. Fragmentation was calculated in the Born
approximation, and the main lacking element is vacuum energy loss
due to gluon radiation caused by the initial hard interaction
\cite{jet-lag}. Such a modification is expected to shrink the
distances $z_2-z_1$ and $z_3-z_1$ and make them $Q^2$-dependent.
Moreover, vacuum energy loss caused by gluon radiation leads to a
distance for pre-hadron production which vanishes in the limit
$z_h\to 1$ \cite{kn,k,knph}. Energy conservation also causes nuclear
suppression toward the kinematical limit $z_h=1$ \cite{knpjs}. These
corrections may only enhance the statements listed above. We plan to
work on this problem and publish elsewhere.

\begin{acknowledgments}

 This work was
supported in part by Fondecyt (Chile) grants, numbers 1050519, 1050589, and
by DFG (Germany)  grant PI182/3-1.

\end{acknowledgments}


 \def\appendix{\par
 \setcounter{section}{0}
\setcounter{subsection}{0}
 \def\thesection{Appendix \Alph{section}}
\def\thesubsection{\Alph{section}.\arabic{subsection}}
\def\theequation{\Alph{section}.\arabic{equation}}
\setcounter{equation}{0}}

 \appendix

\section{Space-time structure of the DIS vertex}\label{avt}

\begin{figure}[htb]
 \includegraphics[width=8cm]{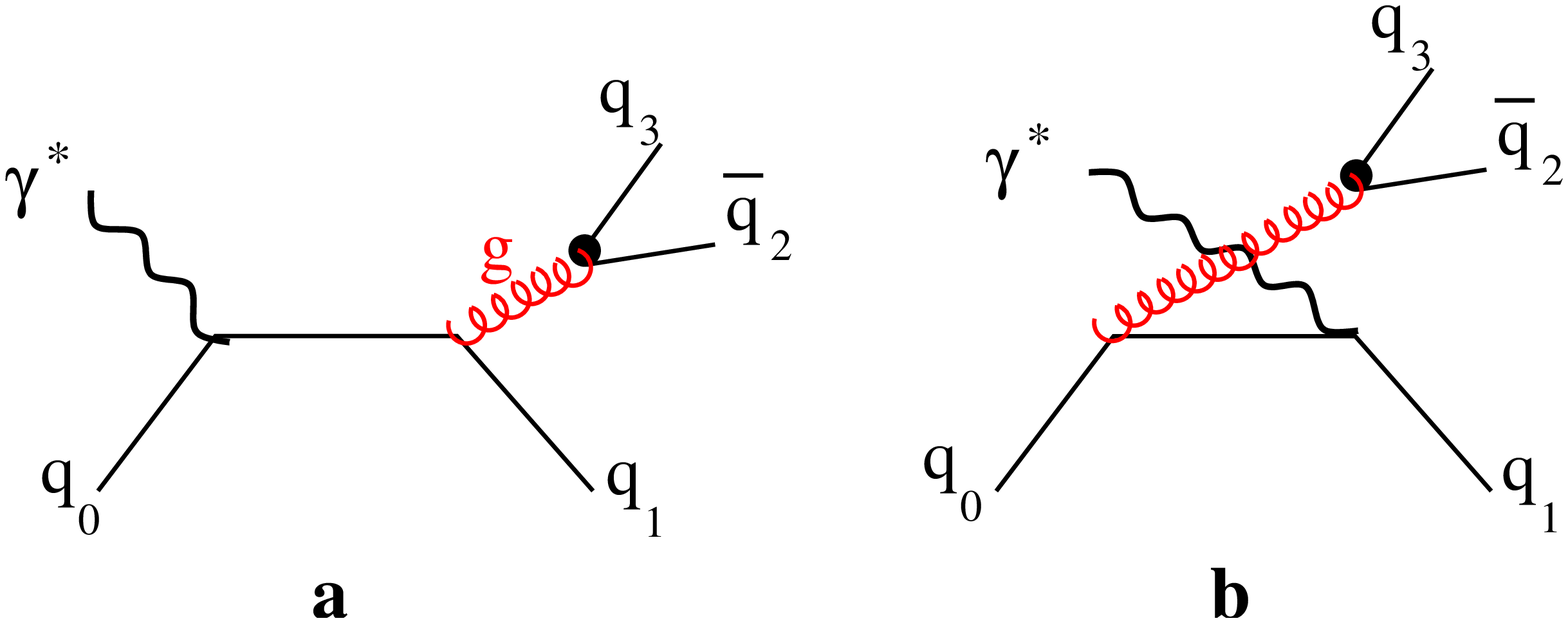}
 \caption{\label{dis} 
 Two Feynman graphs for reaction $\gamma^*q_1\to q_1^\prime\bar q_2q_3$. }
 \end{figure}

The amplitude of the process $l q_0\to l^\prime q_1\bar q_2q_3$
has the form,
\beq
M=\frac{j_\mu^{(l)} J_\mu^{(h)}}{Q^2},
\label{10c}
\eeq
where $j_\mu^{(l)}$ and $J_\mu^{(h)}$ are leptonic and hadronic currents respectively. The latter can be presented as a sum of two terms,
\beq
J_\mu^{(h)}=
J_\mu^{(a)}+
J_\mu^{(b)},
\label{20c}
\eeq
corresponding to graphs Fig.~\ref{dis}a and b, respectively.
The amplitudes have the form,
\beqn
J_\mu^{(a)}&\propto&
\frac{1}{M^2}\,
\bar u(p_1)\gamma_\alpha\hat G(p_1+p_2+p_3)\gamma_\mu u(p_0)
\nonumber\\ &\times&
\bar u(p_3)\gamma_\alpha v(p_2);
\label{25c} \\
J_\mu^{(b)}&\propto&
\frac{1}{M^2}\,
\bar u(p_1)\gamma_\alpha\hat G(p_1-Q)\gamma_\mu
u(p_0)
\nonumber \\ &\times&
\bar u(p_3)\gamma_\alpha v(p_2),
\label{30c}
\eeqn
where $M^2=(p_2+p_3)^2$.

For a massless quark, $m_q=0$, 
\beq
\hat G(p)=\frac{\hat p}{p^2+io}.
\label{40c}
\eeq
Correspondingly,
\beqn
\hat G(x_2-x_1)&=&\frac{1}{(2\pi)^4}
\int d^4p\,\hat G(p)\,
e^{ip(x_2-x_1)} \nonumber\\ &=&
\frac{\gamma_\mu x_\mu}
{2\pi^2[(x_2-x_1)^2-io]^2}.
\label{50c}
\eeqn
The propagator $\hat G(x_2-x_1)$ describes propagation of a quark through the interval $x_2-x_1$ absorbing the virtual photon at one point, and radiating a virtual gluon at another point. 

One can obtain an amplitude of the reaction $lq_0\to l^\prime\pi q_3$ from the amplitude Eq.~(\ref{10c}) projecting the produced $\bar q_2 q_1$ pair to the pion wave function.
Generally, this is quite a nontrivial problem, which, however, much simplifies in some approximations. In the Berger model of a "frozen" pion \cite{berger} one neglects the intrinsic motion of the quarks in the pion,  $p_2=p_1={1\over2}p_\pi$, which is certainly not a realistic approximation. In this case the projection is fulfilled using the relation \cite{born},
\beq
\sum\limits_\lambda \bar u^{(\lambda)}(p_1)\,
v^{(-\lambda)}(p_2)\,{\rm sgn}(\lambda)  =
{1\over2}\gamma_5\left(\hat p_\pi+m_\pi\right).
\label{60c}
\eeq

Another approximation is to neglect the quark and pion masses, $m_q=m_\pi=0$.
In this case,
\beq
\sum\limits_\lambda \bar u^{(\lambda)}(p_1)\,
v^{(-\lambda)}(p_2)\,{\rm sgn}(\lambda)  =
\sqrt{\alpha(1-\alpha)}\gamma_5\,\hat p_\pi,
\label{70c}
\eeq
where $\alpha$ is the fractional light-cone momentum of one of the valence quarks in the pion.
Then the projections of the components of the hadronic current to the final $\pi q_3$ state have 
the form,
\beqn
\bar J_\mu^{(a)}&=&\int\limits_0^1 d\alpha\,
\frac{\sqrt{2}\phi_\pi(\alpha)}{\bar M^2}\,
\bar u(p_3)\gamma_5\hat p_\pi \hat G(p_3+p_\pi)
\gamma_\mu u(p_0);\nonumber\\
\label{75c}\\
\bar J_\mu^{(b)}&=&\int\limits_0^1 d\alpha\,
\frac{\sqrt{2}\phi_\pi(\alpha)}{\bar M^2}\,
\bar u(p_3)\gamma_5\hat G(\alpha p_\pi-Q)\gamma_\mu
\hat p_\pi u(p_0),
\nonumber\\
\label{80c}
\eeqn
where $\bar M^2=\bigl[p_3+(1-\alpha)p_\pi\bigr]^2$.

Since according to the Dirac equation, $\bar u(p_3)\hat p_3=0$, we can replace the product $\hat p_\pi\hat G(p_\pi+p_3)$ in the component $\bar J_\mu^{(a)}$, Eq.~(\ref{75c}), by
\beq
G(p_\pi+p_3)=(\hat p_\pi+\hat p_3)
\hat G(p_\pi+p_3)\equiv 1.
\label{90c}
\eeq
Then the effective propagator,
\beq
G(x_2-x_1)=\frac{1}{(2\pi)^4}
\int d^4p\,G(p)\,e^{ip(x_2-x_1)}=
\delta(x_2-x_1),
\label{100c}
\eeq
is not zero only when $x_2=x_1$, i.e. the virtual gluon is radiated at the same point where the virtual photon is absorbed.

For the second component of the hadronic current, $\bar J_\mu^{(b)}$, Eq.~(\ref{80c}),
it was demonstrated in \cite{born} that its longitudinal-to-transverse contribution ratio is suppressed as 
$\sigma_L/\sigma_T\sim p_{\pi\perp}^2/(1-z)Q^2$. Therefore, we will keep only the transverse part of the current $\bar J_T^{(b)}$.

Then, using the kinematic relation, $p_\pi=z_hp_3/(1-z_h)+O(p_{\pi\perp})$ and applying the Dirac equation, $\bar u(p_3)\hat p_3=0$, the transverse part of the hadronic current Eq.~(\ref{80c}) can be modified as,
\beqn
\bar u(p_3)\gamma_5\hat G(\alpha p_\pi-Q)\gamma_\mu
\hat p_\pi u(p_0) &=&
\tilde G\, \bar u(p_3)\gamma_5\gamma_\perp u(p_0)
\nonumber\\ &+&
O\left(\frac{p_{\pi\perp}^2}{Q^2}\right),
\label{110c}
\eeqn
where
\beq
\tilde G = \frac{z_h^2 Q^2+p_{\pi\perp}^2}
{Q^2 z_h(1-\alpha z_h)-\alpha p_{\pi\perp}^2} =
\frac{z_h}{1-\alpha z_h}+
O\left(\frac{p_{\pi\perp}^2}{Q^2}\right).
\label{120c}
\eeq
Thus, neglecting corrections of the order of $\la p_{\pi\perp}^2\ra/Q^2$ we arrive at the relation between the two components of the hadronic current,
\beq
\bar J_\mu^{(b)}=\frac{z_h}{1-\alpha z_h}\,
\bar J_\mu^{(a)}.
\label{130c}
\eeq
Therefore, the effective quark propagator for the second component of the hadronic current, corresponding to Fig.~\ref{dis}, should have the same property Eq.~(\ref{100c}) shrinking the interval $x_2-x_1$.

\section{Multi-parton cross sections \boldmath$\Sigma_i$}
\label{sigmas} \setcounter{equation}{0}

The effective cross sections $\Sigma_i$, which are linear
combinations of dipole $\bar qq$ cross sections, can be derived
within the Born approximation. The derivation is quite lengthy and
not easy. It is much easier to used a set of equation which
correspond to different limiting configurations within the
multi-parton state. The way how it works is explained further in
concrete examples.

\subsection{4-body cross section
\boldmath$\Sigma_1(\vec r^{\,q_1},\vec s^{\,q_1},\vec r^{\,g},\vec s^{\,g})$}\label{sigma4}

To proceed, let us start with $\Sigma_1$, which is the total cross
section of a 4-parton colorless system $gq_1\,\bar g\bar q_1$, in
which the $gq_1$ and $\bar g\bar q_1$ pairs are in color triplet and
anti-triplet states respectively, while the pairs $\bar gg$ and
$\bar q_1q_1$ are color singlets. Apparently, the system is
symmetric relative to interchanges $\verg\leftrightharpoons\vesg$
and $\verq\leftrightharpoons\vesq$, so we get,
\beqn &&\Sigma_1(\vec
r^{\,q_1},\vec s^{\,q_1},\vec r^{\,g},\vec s^{\,g})=
a_1\bigl[\sq(\verg-\verq) + \sq(\vesg-\vesq)\bigr] \nonumber\\ &+&
b_1\bigl[\sq(\verg-\vesq) + \sq(\vesg-\verq)\bigr] +
c_1\sq(\verq-\vesq) \nonumber\\ &+& d_1\sq(\verg-\vesg).
\label{a100}
\eeqn

This expression, at $\vesg=\verq$, must turn into the known 3-body
($g, q, \bar q$) cross section,
\beqn \Sigma_1\Rightarrow
\sigma_3(\verg,\vesq,\verq)&=&{9\over8}\bigl[\sq(\verg-\verq)+
\sq(\verg-\vesq)\bigr] \nonumber\\ &-& {1\over8}\sq(\verq-\vesq),
\label{a200}
\eeqn
leading to the relations, $a_1=b_1+d_1={9\over8}$
and $b_1+c_1=-{1\over8}$. Otherwise, fixing $\verq=\vesq$, we
expect,
\beq \Sigma_1\Rightarrow
\sg(\verg-\vesg)={9\over4}\,\sq(\verg-\vesg). \label{a300}
\eeq
This leads to new relations, $a_1+b_1=0$ and $d_1={9\over4}$.

Thus, we arrive at the coefficients in (\ref{a100}), $a_1=-b_1={9\over8}$,
$c_1=1$ and $d_1={9\over4}$, which proves Eq.~(\ref{1000}).

\subsection{Cross sections \boldmath$\Sigma_2(\verqc,\verqb,\verqa,
\vesg,\vesqa)$ and $\Sigma_3(\verqc,\verqb,\verqa,\vesqc,\vesqb,\vesqa)$}

Again, relying on the symmetry relative the replacement $\vec
r\leftrightharpoons\vec s$ we can write, \beqn \Sigma_3 &=&
a_3\bigl[\sq(\verqa-\verqb) + \sq(\vesqa-\vesqb)\bigr] \nonumber\\
&+& b_3\bigl[\sq(\verqa-\verqc) + \sq(\vesqa-\vesqc)\bigr]
\nonumber\\ &+& c_3\bigl[\sq(\verqb-\verqc) +
\sq(\vesqb-\vesqc)\bigr] \nonumber\\ &+& d_3\bigl[\sq(\verqa-\vesqb)
+ \sq(\vesqa-\verqb)\bigr] \nonumber\\ &+&
e_3\bigl[\sq(\verqa-\vesqc) + \sq(\vesqa-\verqc)\bigr] \nonumber\\
&+& f_3\bigl[\sq(\verqb-\vesqc) + \sq(\vesqb-\verqc)\bigr]
\nonumber\\ &+& g_3\sq(\verqa-\vesqa) + h_3\sq(\verqb-\vesqb)
\nonumber\\ &+& k_3\sq(\verqc-\vesqc) \label{a400}
\eeqn
We can simplify this expression by considering known limiting combinations.
For $\verqb=\verqc\equiv\verg$ and $\vesqb=\vesqc\equiv\vesg$ we
have to get
$\Sigma_3\Rightarrow\Sigma_1(\verg,\verqa,\vesg,\vesqa)$. This
condition leads to the following relations, $a_3+b_3=-d_3-e_3
={9\over8}$, $g_3=-1$, $2f_3+h_3+k_3={9\over4}$.

The next possibility is to fix $\vesqb=\vesqc\equiv\vesg$ and
$\verqa=\vesqa$. Then we should arrive at the 3-body case,
\beqn
\Sigma_3&\Rightarrow& \sigma_3(\verqb,\verqc,\vesg) =
{9\over8}\bigl[\sq(\vesg-\verqb) \nonumber\\ &+& \sq(\vesg-\verqc)-
{1\over8}\,\sq(\verqb-\verqc). \label{a500}
\eeqn
This results in additional relations, $f_3+h_3=f_3+k_3={9\over8}$, $c_3=-{1\over8}$
and $a_3+d_3=b_3+c_3=0$.

Eventually, after fixing the coordinates differently,
$\vesqb=\vesqc\equiv\vesg$ and $\verqa=\verqb$, Eq.~(\ref{a400})
simplifies to,
\beq
\Sigma_3\Rightarrow\sigma_3(\vesqb,\verqc,\vesg). \label{a600}
\eeq
 Correspondingly, new relations emerge, $a_3+b_3=k_3+f_3={9\over8}$,
 $b_3+c_3=d_3+g_3=h_3+f_3+d_3+e_3=0$, and $e_3=-{1\over8}$.

Solving these sets of relations we get the coefficients in (\ref{a400}),
\beqn
a_3&=&-d_3=g_3=h_3=k_3=1,
\nonumber\\
b_3&=&-e_3=-c_3=f_3={1\over8},
\label{a700}
\eeqn
which lead to Eq.~(\ref{2500}).

Eventually, one can get Eq.~(\ref{2400}) for $\Sigma_2(\verqc,\verqb,\verqa,
\vesg,\vesqa)$ by fixing $\vesqb=\vesqc\equiv\vesg$ in
Eq.~(\ref{2500}) for $\Sigma_3(\verqc,\verqb,\verqa,
\vesqc,\vesqb,\vesqa)$.

\section{Functions \boldmath$W_i$}
\label{W} \setcounter{equation}{0}

\subsection{Full calculation}

As we already mentioned, equations (\ref{800})-(\ref{850}) and
(\ref{1800})-(\ref{2100}) can be solved analytically, provided that
the nuclear density is constant, $\rho_A(b,z)=\rho_0$, and the
dipole cross section has the simple form $\sq=C\,r^2$. In this case
the equations, which are bilinear in the interaction potential, can
be solved following Ref.~\cite{feynman}.

We demonstrate here the method for the example of Eq.~(\ref{800}),
which can be represented as,
\beq i\,\frac{\partial W_1}{\partial
z_2} = H_1\,W_1. \label{100b}
\eeq
The effective Hamiltonian $H_1$
can be written as a sum of the effective kinetic and potential
energies,
\beq H_1=T_1+V_1. \label{200b}
\eeq T
hen $W_1$ can be presented as,
\beq W_1(z_2,z_1)=
N_1(z_2,z_1)\,e^{iS_1},
\label{300b}
\eeq
where
\beq
S_1=\int\limits_{z_1}^{z_2} dz\,L_1(z).
\label{400b}
\eeq
The effective Lagrangian has the form,
\beq
L_1=T_1(z)-V_1(z),
\label{500b}
\eeq
where the potential term reads
\beq
V_1(z)=-i\rho_0\Sigma_1\bigl(\verg(z),\verqa(z),\vesg(z),\vesqa(z)\bigr).
\label{550b}
\eeq

The kinetic term has the form,
\beqn
T_1(z)&=&
{1\over2}E_g\bigl(
\vec v^{\,g}(z)\bigr)^2
+
{1\over2}E_1\bigl(\vec v^{\,q_1}(z)\bigr)^2
\nonumber\\ &-&
{1\over2}\bar E_g\bigl(\vec u^{\, g}(z)\bigr)^2
-
{1\over2}\bar E_1\bigl(\vec u^{\,q_1}(z)\bigr)^2
,
\label{600b}
\eeqn
where
\beqn
\vec v^{\,g}(z)&=&{d\over dz}\verg(z);\nonumber\\
\vec v^{\,q_1}(z)&=&{d\over dz}\verqa(z);\nonumber\\
\vec u^{\,g}(z)&=&{d\over dz}\vesg(z);\nonumber\\
\vec u^{\,q_1}(z)&=&{d\over dz}\verqa(z).
\label{700b}
\eeqn
The transverse separations as functions of $z$ are the solutions of the Euler-Lagrange
differential equations,
\beqn
{d\over dz}\left(\frac{\partial L_1}{\partial \vec v^{\,g}}\right) -
\frac{\partial L_1}{\partial \verg} &=& 0;\nonumber\\
{d\over dz}\left(\frac{\partial L_1}{\partial \vec v^{\,q_1}}\right) -
\frac{\partial L_1}{\partial \verqa} &=& 0;\nonumber\\
{d\over dz}\left(\frac{\partial L_1}{\partial \vec u^{\,g}}\right) -
\frac{\partial L_1}{\partial \vesg} &=& 0;\nonumber\\
{d\over dz}\left(\frac{\partial L_1}{\partial \vec u^{\,q_1}}\right) -
\frac{\partial L_1}{\partial \vesqa} &=& 0;
\label{800b}
\eeqn
It is convenient to use the coordinates of the center of mass,
\beq
\vec R=\frac{E_g\verg+E_1\verqa}
{E_g+E_1}=
(1-x)\verg+x\verqa,
\label{900b}
\eeq
where $x=\alpha z_h$, and
\beq
\vec S=\frac{\bar E_g\vesg+\bar E_1\vesqa}
{\bar E_g+\bar E_1}=
(1-\bar x)\vesg+\bar x\vesqa,
\label{1000b}
\eeq
where $\bar x=\bar\alpha z_h$.

The relative separations are given by,
\beqn
\vec r&=&\verg-\verqa;\nonumber\\
\vec s &=& \vesg-\vesqa.
\label{1100b}
\eeqn

The corresponding velocities read,
\beqn
\vec V(z)&=&{d\over dz}\,\vec R(z);\nonumber\\
\vec U(z)&=&{d\over dz}\,\vec S(z);\nonumber\\
\vec v(z)&=&{d\over dz}\,\vec r(z);\nonumber\\
\vec u(z)&=&{d\over dz}\,\vec s(z).
\label{1200b}
\eeqn

In the new variables the Lagrangian Eq.~(\ref{500b}) gets the form,
\beqn
&&L_1(z)=
{1\over2}E\left(\vec V^{\,2}-\vec U^{\,2}\right)
+{1\over2}\varepsilon\,\vec v^{\,2}
\nonumber\\ &-&
{1\over2}\bar\varepsilon\,\vec u^{\,2}+
{i\over2}\rho_0\Sigma_1(\vec r,\vec s,\vec R-\vec S),
\label{1300b}
\eeqn
where $E=E_g+E_1=\bar E_g+\bar E_1=E$.

Then, we make the following combinations of the centers of gravity coordinates,
\beqn
\vec R^{\,+}&=&{1\over2}(\vec R+\vec S);
\nonumber\\
\vec R^{\,-}&=&\vec R-\vec S;
\label{1400b}
\eeqn
and velocities,
\beqn
\vec V^{\,+}&=&{1\over2}(\vec V+\vec U);
\nonumber\\
\vec V^{\,-}&=&\vec V-\vec U;
\label{1500b}
\eeqn

Notice that the cross section $\Sigma_1$ Eq.~(\ref{1000}), which
enters the potential term of the Lagrangian, Eq.~(\ref{550b}), is
independent of $R^+$. Therefore, the Euler-Lagrange equations
(\ref{800b}) written via new variables $\vec R^{\,+}, \vec R^{\,-},\
\vec r,\ \vec s$ have a simple solution, ${d\over dz}\,\vec
V^{\,-}=0$, i.e. \beqn \vec V^{\,-}(z)&=& Const = \frac{\vec
R_2^{\,-}-\vec R_1^{\,-}} {z_2-z_1};
\nonumber\\
\vec R^{\,-}(z)&=&\vec R_1^{\,-}+(z-z_1)\vec V^{\,-}.
\label{1600b}
\eeqn

Then, for the first term in the Lagrangian Eq.~(\ref{1300b}) the
integral Eq.~(\ref{400b}) can be calculated as,
\beqn &&
{E\over2}\int\limits_{z_1}^{z_2} dz(\vec V^{\,2}-\vec U^{\,2})=
E\int\limits_{z_1}^{z_2} dz \,\vec V^{\,+}\vec V^{\,-} \nonumber\\
&=& \frac{E}{z_2-z_1}\, (\vec R_2^{\,+}-\vec R_1^{\,+}) (\vec
R_2^{\,-}-\vec R_1^{\,-}). \label{1700b}
\eeqn
Thus, for this part
of the integral Eq.~(\ref{400b}) we did not need to know the
explicit form of $\vec R^{\,+}(z)$, which is rather complicated.

In order to calculate the rest of the integral (\ref{400b}), we need
to know $\vec r(z)$ and $\vec s(z)$. The potential Eq.~(\ref{550b})
can be represented as,
\beqn V_1&=&{1\over2}\biggl[a\vec
r^{\,2}-2b\,\vec r\cdot\vec s+ c\vec s^{\,2}+ 2d\,\vec r\cdot\vec
R^{\,-} \nonumber\\ &-&
 2e\,\vec s\cdot\vec R^{\,-}+
f\left(\vec R^{\,-}\right)^2\biggr],
\label{1800b}
\eeqn
where
\beqn
a&=&-iC\rho_0\left[1+\alpha^2z_h^2+{\alpha z_h}{4}\right];
\nonumber\\
b&=&-iC\rho_0\left[1+\alpha\bar\alpha z_h^2+(\alpha+\bar\alpha)
\frac{z_h}{4}\right];
\nonumber\\
c&=&-iC\rho_0\left[1+\bar\alpha^2z_h^2+{\bar\alpha z_h}{4}\right];
\nonumber\\
d&=&-iC\rho_0\left[\alpha z_h+{1\over 8}\right];
\nonumber\\
e&=&-iC\rho_0\left[\bar\alpha z_h+{1\over 8}\right];
\nonumber\\
f&=&-iC\rho_0
\label{1850b}
\eeqn

Then the Euler-Lagrange equations lead to the following linear equations for
$\vec r(z)$ and $\vec s(z)$,
\beqn
\varepsilon\,\left(\frac{d}{dz}\right)^2\vec r&=&
-a\vec r+b\vec s-d\vec R^{\,-};
\nonumber\\
\bar\varepsilon\,\left(\frac{d}{dz}\right)^2\vec s&=&
-b\vec r+c\vec s-e\vec R^{\,-},
\label{1900b}
\eeqn
where $\varepsilon=x(1-x)E$, $\bar\varepsilon=\bar x(1-\bar x)E$.
To make these equations homogeneous we switch to new variables,
\beqn
\vec r^{\,\prime}&=&\vec r+\upsilon \vec R^{\,-};
\nonumber\\
\vec s^{\,\prime}&=&\vec s+\zeta \vec R^{\,-},
\label{2000b}
\eeqn
where $\upsilon$ and $\zeta$ are solutions of the algebraic equations,
\beqn
\upsilon a-\zeta b&=&d
\nonumber\\
\upsilon b-\zeta c&=&e.
\label{2100b}
\eeqn
Then, $\vec r^{\,\prime}(z)$ and $\vec s^{\,\prime}(z)$ satisfy the homogeneous equations,
\beqn
\varepsilon\,\frac{d^2\vec r^{\,\prime}}{dz^2}&=&
-a\vec r^{\,\prime}+b\vec s^{\,\prime};
\nonumber\\
\bar\varepsilon\,\frac{d^2\vec s^{\,\prime}}{dz^2}&=&
-b\vec r^{\,\prime}+c\vec s^{\,\prime}.
\label{2200b}
\eeqn
The solution of these equation is,
\beqn
\vec r^{\,\prime}(z)&=&\vec A\,\sin(\omega_1z)+
\vec B\,\cos(\omega_1z)\nonumber\\ &+&
\nu\bigl[\vec C\,\sin(\omega_2z)+
\vec D\,\cos(\omega_2z)\bigr];
\label{2300b}\\
\vec s^{\,\prime}(z)&=& \mu\bigl[\vec A\,\sin(\omega_1z)+ \vec
B\,\cos(\omega_1z)\bigr] \nonumber\\ &+& \vec C\,\sin(\omega_2z)+
\vec D\,\cos(\omega_2z). \label{2350b}
\eeqn
Here $\omega_{1,2}=\sqrt{\lambda_{1,2}}$, where $\lambda_{1,2}$ are the
solutions of the quadratic equation
$(a-\epsilon\lambda)(c+\tilde\epsilon\lambda)-b^2=0$, and
$\mu=(a-\epsilon\lambda_1)/b$, $\nu=(c+\tilde\epsilon\lambda_2)/b$.

The vectors $\vec A,\ \vec B,\ \vec C,\ \vec D$ are fixed by the boundary conditions
$\vec r^{\,\prime}(z_{1,2})=\vec r^{\,\prime}_{1,2}$ and
$\vec s^{\,\prime}(z_{1,2})=\vec s^{\,\prime}_{1,2}$,
\beqn
\vec A&=& \frac{1}{1-\mu\nu}\,
\frac{\vec\rho_1 c_{12}-\vec\rho_2 c_{11}}
{s_{11}c_{12}-s_{12}c_{11}};
\nonumber\\
\vec B&=& \frac{1}{1-\mu\nu}\,
\frac{\vec\rho_1 s_{12}-\vec\rho_2 s_{11}}
{c_{11}s_{12}-c_{12}s_{11}};
\nonumber\\
\vec C&=& \frac{1}{1-\mu\nu}\,
\frac{\vec\tau_1 c_{22}-\vec\tau_2 c_{21}}
{s_{21}c_{22}-s_{22}c_{21}};
\nonumber\\
\vec D&=& \frac{1}{1-\mu\nu}\,
\frac{\vec\tau_1 s_{22}-\vec\tau_2 s_{21}}
{c_{21}s_{22}-c_{22}s_{21}},
\label{2400b}
\eeqn
where
\beqn
\vec\rho_{1,2}(z)&=&\vec r^{\,\prime}_{1,2}(z)-
\mu\,\vec s^{\,\prime}_{1,2}(z);
\nonumber\\
\vec\tau_{1,2}(z)&=&\vec s^{\,\prime}_{1,2}(z)-
\nu\,\vec r^{\,\prime}_{1,2}(z);
\label{2500b}
\eeqn
and $s_{i,j}=\sin(\omega_iz_j)$, $c_{i,j}=\cos(\omega_iz_j)$, $i,j=1,2$.

Now we are in a position to perform the rest of integration in Eq.~(\ref{400b}),
and we arrive at the final expression for the action,
\begin{widetext}
\beqn
S_1&=&\frac{E}{\Delta z_{12}}\,
(\vec R_2^{\,+}-\vec R_1^{\,+})
(\vec R_2^{\,-}-\vec R_1^{\,-}) +
\frac{\omega_1Ex(1-x)}{2(1-\mu\nu)}\,
\left[(\vec\rho_1^{\,2}+\vec\rho_2^{\,2})\cot(\omega_1\Delta z_{12})-
\frac{2\vec\rho_1\cdot\vec\rho_2}{\sin(\omega_1\Delta z)}\right]
\nonumber\\ &-&
\frac{\omega_2E\bar x(1-\bar x)}{2(1-\mu\nu)}\,
\left[(\vec\tau_1^{\,2}+\vec\tau_2^{\,2})\cot(\omega_2\Delta z_{12})-
\frac{2\vec\tau_1\cdot\vec\tau_2}{\sin(\omega_2\Delta z_{12})}\right]-
\frac{Ex(1-x)v}{\Delta z_{12})}\,(\vec r_2-\vec r_1)\left(\vec R_2^{\,-}-\vec R_1^{\,-}\right)
\nonumber\\ &+&
\frac{E\bar x(1-\bar x)\zeta}{\Delta z_{12}}\,(\vec s_2-\vec s_1)\left(\vec R_2^{\,-}-\vec
R_1^{\,-}\right)
+\frac{(\vec R_2^{\,-}-\vec R_1^{\,-})^2}{2\Delta z_{12}}\
\biggl[-Ex(1-x)v^2+
E\bar x(1-\bar x)\zeta^2\biggr]
\nonumber\\ &-&
{1\over6}\bigl(f+v^2a+\zeta^2c-2v\zeta b\bigr)
\left[\left(\vec R_2^{\,-}\right)^2+
\vec R_2^{\,-}\cdot\vec R_1^{\,-}+
\left(\vec R_1^{\,-}\right)^2\right]\,\Delta z_{12}.
\label{2600b}
\eeqn
\end{widetext}
where $\Delta z_{12}=z_2-z_1$; $x=\alpha z_h$; $\bar x=\bar\alpha z_h$.

Now we can solve equation~(\ref{100b}),
\beq
\left(i\,\frac{\partial}{\partial z_2}-H_1\right)\,
e^{iS_1}=i\Phi(\Delta z_{12})\,e^{iS_1},
\label{2700b}
\eeq
where
\beq
\Phi(\Delta z_{12})=\frac{2}{\Delta z_{12}}+\omega_1\cot(\omega_1\Delta z_{12})
+\omega_2\cot(\omega_2\Delta z_{12}).
\label{2800b}
\eeq
Then, from (\ref{800}) and the boundary condition (\ref{900}) we find the factor in
(\ref{300b}),
\beq
N_1(\Delta z_{12})=\left(\frac{E}{2\pi\Delta z_{12}}\right)^2\,
\frac{E^2x(1-x)\bar x(1-\bar x)\omega_1\omega_2}
{(2\pi)^2\sin(\omega_1\Delta z_{12})\sin(\omega_2\Delta z_{12})}.
\label{2900b}
\eeq

The derivation of the functions $W_{2,3}(z_2,z_1)$ is analogous, but rather cumbersome,
so we skip it here.

\subsection{Approximations}

\subsubsection{Function $W_1$}

The expressions for $W_{i}(z_2,z_1)$ ($i=1,\ 2,\ 3$) simplify, if
$x=\bar x$ (i.e. $\alpha=\bar\alpha$), and the parameters $a,...\ f$ in
Eq.~(\ref{1800b}), which are functions of $x$ and $\bar x$, are
related if $x=\bar x$,
\beqn
a(x=\bar x)&=& b(x=\bar x)= c(x=\bar x)\nonumber\\
d(x=\bar x)&=&e(x=\bar x) \label{3000b}
\eeqn
Besides, for the
parameters defined in (\ref{1900b}) $\varepsilon=\bar\varepsilon$, and
the parameters $\mu,\nu\to 1$. It turns out that it is more
complicated to perform a transition in the found solution for the
action $S_1(\alpha-\bar\alpha\to0)$, than to repeat the derivation
specifically in this limit.

In this case the Lagrangian (\ref{500b}) gets a simple form,
\beqn
&&
L_1(\alpha=\bar\alpha)=E\,\vec V^{\,+}\cdot\vec V^{\,-}+
Ex(1-x)\vec v^{\,+}\cdot\vec v^{\,-}
\nonumber\\&-&
{1\over2}\bigl[
a(\vec r^{\,-})^2+f(\vec R^{\,-})^2+
2e\vec r^{\,-}\vec R^{\,-}\bigr],
\label{3100b}
\eeqn
where $\vec v^{\,\pm}=\frac{\partial}{\partial z}\vec r^{\,\pm}$,
$\vec r^{\,+}=(\vec r+\vec s)/2$, $\vec r^{\,-}=\vec r-\vec s$.

From the Euler-Lagrange equations of motion it follows that
$\frac{\partial}{\partial z}\vec V^{\,-}=\frac{\partial}{\partial z}\vec v^{\,-}=0$, so
\beqn
\vec V^{\,-}&=&\frac{\vec R_2^{\,-}-\vec R_1^{\,-}}{\Delta z_{12}};
\nonumber\\
\vec v^{\,-}&=&\frac{\vec r_2^{\,-}-\vec r_1^{\,-}}{\Delta z_{12}};
\label{3150b}
\eeqn
and
\beqn
\vec R^{\,-}&=&\vec R_1^{\,-}+\vec v^{\,-}(z-z_1);
\nonumber\\
\vec r^{\,-}&=&\vec r_1^{\,-}+\vec v^{\,-}(z-z_1).
\label{3250b}
\eeqn

This is sufficient for calculating the action Eq.~(\ref{400b}), and we arrive at,

\begin{widetext}

\beqn
S_1&=&{E\over\Delta z_{12}}\left[\left(
\vec R_2^{\,+}-\vec R_1^{\,+}\right)
\left(\vec R_2^{\,-}-\vec R_1^{\,-}\right)
+x(1-x)\left(
\vec r_2^{\,+}-\vec r_1^{\,+}\right)
\left(\vec r_2^{\,-}-\vec r_1^{\,-}\right)\right]-
\frac{a\Delta z_{12}}{6}
\left[\left(\vec r_2^{\,-}\right)^2+
\vec r_2^{\,-}\vec r_1^{\,-}
+\left(\vec r_1^{\,-}\right)^2\right]
\nonumber\\ &-&
\frac{f\Delta z_{12}}{6}
\left[\left(\vec R_2^{\,-}\right)^2+
\vec R_2^{\,-}\vec R_1^{\,-}
+\left(\vec R_1^{\,-}\right)^2\right]-
\frac{e\Delta z_{12}}{6}
\left[2\vec R_2^{\,-}\vec r_2^{\,-}+
2\vec R_1^{\,-}\vec r_1^{\,-}+
\vec R_2^{\,-}\vec r_1^{\,-}+
\vec R_1^{\,-}\vec r_2^{\,-}
\right]
\label{3350b}
\eeqn
\end{widetext}

Then the coefficient $N_1(\Delta z_{12})$ in Eq.~(\ref{300b}) gets the very simple form,
\beq
N_1(\Delta z_{12})=x(1-x)\left(\frac{E}{2\pi\Delta z_{12}}\right)^4.
\label{3450b}
\eeq

\subsubsection{Function $W_3$}

The next case is $W_3$, which is simple due to the symmetry relative to interchange
$\vec r^{\,q_i}\Leftrightarrow \vec s^{\,q_i}$. First we introduce the Jacoby coordinates,
\beqn
\vec R &=& x_1\verqa+x_2\verqb+x_3\verqc;
\nonumber\\
\vec r &=& \frac{x_2\verqb+x_3\verqc}
{x_2+x_3} - \verqa;
\nonumber\\
\vec\rho &=& \verqc-\verqb;
\nonumber\\
\vec S&=&x_1\vesqa+x_2\vesqb+x_3\vesqc;
\nonumber\\
\vec s &=& \frac{x_2\vesqb+x_3\vesqc}
{x_2+x_3} - \vesqa;
\nonumber\\
\vec\tau &=& \vesqc-\vesqb.
\label{3550b}
\eeqn
Here $x_1=\alpha z_h$, $x_2=(1-\alpha)z_h$, $x_3=1-z_h$.

We also introduce combinations of the Jacoby coordinates, $\vec
R^{\,+}={1\over2}(\vec R+\vec S)$, $\vec R^{\,-}=\vec R-\vec S$,
$\vec r^{\,+}={1\over2}(\vec r+\vec s)$, $\vec r^{\,-}=\vec r-\vec
s$, $\vec\rho^{\,+}={1\over2}(\vec\rho+\vec\tau)$,
$\vec\rho^{\,-}=\vec\rho-\vec\tau$.

The Lagrangian $L_3$ can be represented as,
\beqn
L_3 &=& E\,\vec V^{\,+}\vec V^{\,-} +
x(1-x)E\,\vec v^{\,+}\vec v^{\,-}+
\mathcal{E}\,\vec \omega^{\,+}\vec\omega^{\,-}
\nonumber\\ &-&
{1\over2}\biggl\{
a\left(\vec r^{\,-}\right)^2+
b\left(\vec\rho^{\,-}\right)^2+
c\left(\vec R^{\,-}\right)^2
\nonumber\\ &+&
2d\,\vec r^{\,-}\vec \rho^{\,-}+
2e\,\vec r^{\,-}\vec R^{\,-}+
2f\,\vec \rho^{\,-}\vec R^{\,-}
\biggr\},
\label{2650b}
\eeqn
where $\vec V^{\,\pm}=\frac{\partial}{\partial z}\vec R^{\,\pm}$,
$\vec v^{\,\pm}=\frac{\partial}{\partial z}\vec r^{\,\pm}$,
$\vec \omega^{\,\pm}=\frac{\partial}{\partial z}\vec \rho^{\,\pm}$,
\beqn
a&=&-iC\rho_0\left[1+\alpha^2z_h^2+\frac{\alpha z_h}{4}\right];
\nonumber\\
b&=&-iC\rho_0\left[1-{9\over4}\frac{(1-\alpha)z_h(1-z_h)}
{1-\alpha _h}\right];
\nonumber\\
c&=&-iC\rho_0;
\nonumber\\
d&=&-iC\rho_0\left[\alpha z_h+{1\over 8}
-{9\over8}\frac{(1-z_h)(1+\alpha z_h)}{1-\alpha z_h}\right];
\nonumber\\
e&=&-iC\rho_0\left[\alpha z_h+{1\over 8}\right];
\label{1870b}
\eeqn

Again, the equations of motion lead to the relations,
\beqn
\frac{\partial}{\partial z}\,\vec V^{\,-} &=&
\frac{\partial}{\partial z}\,\vec v^{\,-}=
\frac{\partial}{\partial z}\,\vec v^{\,-}=0;
\nonumber\\
\vec R^{\,-} &=& \vec R_1^{\,-} + (\vec R_1^{\,-}-\vec R_2^{\,-})
\,\frac{z-z_3}{\Delta z_{34}};
\nonumber\\
\vec r^{\,-} &=& \vec r_1^{\,-} + (\vec R_1^{\,-}-\vec r_2^{\,-})
\,\frac{z-z_3}{\Delta z_{34}};
\nonumber\\
\vec \rho^{\,-} &=& \vec \rho_1^{\,-} + (\vec \rho_1^{\,-}-\vec \rho_2^{\,-})
\,\frac{z-z_3}{\Delta z_{34}}.
\label{2750b}
\eeqn
These relations lead to the following action,
\begin{widetext}

\beqn
S_3&=&
\int\limits_{z_3}^{z_4} dz\,L_3
\nonumber\\ &=&
\frac{1}{\Delta z_{34}}\left[
E\left(\vec R_2^{\,+}-\vec R_1^{\,+}\right)
\left(\vec R_2^{\,-}-\vec R_1^{\,-}\right)+
E\,x(1-x)\left(\vec r_2^{\,+}-\vec r_1^{\,+}\right)
\left(\vec r_2^{\,-}-\vec r_1^{\,-}\right)+
\varepsilon\left(\vec \rho_2^{\,+}-\vec \rho_1^{\,+}\right)
\left(\vec \rho_2^{\,-}-\vec \rho_1^{\,-}\right)\right]
\nonumber\\ &-& {\Delta z_{34}\over6}\Biggl\{
a\left[\left(\vec r_2^{\,-}\right)^2+
+\vec r_2^{\,-}\vec r_1^{\,-}+
\left(\vec r_1^{\,-}\right)^2\right]+
b\left[\left(\vec \rho_2^{\,-}\right)^2+
+\vec \rho_2^{\,-}\vec \rho_1^{\,-}+
\left(\vec \rho_1^{\,-}\right)^2\right]+
c\left[\left(\vec R_2^{\,-}\right)^2+
+\vec R_2^{\,-}\vec R_1^{\,-}+
\left(\vec R_1^{\,-}\right)^2\right]
\nonumber\\ &+&
d\left[
2\vec r_2^{\,-}\vec \rho_2^{\,-}+2\vec r_1^{\,-}\vec \rho_2^{\,-}+
\vec r_2^{\,-}\vec \rho_1^{\,-}+\vec r_2^{\,-}\vec \rho_2^{\,-}\right]+
e\left[2\vec r_2^{\,-}\vec R_2^{\,-}+2\vec r_1^{\,-}\vec R_2^{\,-}+
\vec r_2^{\,-}\vec R_1^{\,-}+\vec r_2^{\,-}\vec R_2^{\,-}\right]
\nonumber\\ &+&
f\left[2\vec \rho_2^{\,-}\vec R_2^{\,-}+2\vec \rho_1^{\,-}\vec R_2^{\,-}+
\vec \rho_2^{\,-}\vec R_1^{\,-}+\vec \rho_2^{\,-}\vec R_2^{\,-}\right]\Biggr\}
\label{2850b}
\eeqn
\end{widetext}
Eventually, we arrive at,
\beq
W_3(z)=N_3(\Delta z_{34})\,e^{iS_3},
\label{2950b}
\eeq
where
\beqn
N_3(\Delta z_{34})&=&\alpha(1-\alpha)z_h^2(1-z_h)
\left(\frac{E}{2\pi\Delta z_{34}}\right)^6
\nonumber\\ &=&
x_1x_2x_3\left(\frac{E}{2\pi\Delta z_{34}}\right)^6.
\label{3050b}
\eeqn

\subsubsection{Function $W_2$}

The calculation of $W_2$ is more involved. First we switch to new
variables.
\beqn
\vec R&=&x_1\verqa + x_2\verqb+x_3\verqc;
\nonumber\\
\vec r&=&\frac{x_2\verqb+x_3\verqc}{x_2+x_3} - \verqa;
\nonumber\\
\vec\rho&=&\verqc-\verqb;
\nonumber\\
\vec S&=&x\vesqa+(1-x)\vesg;
\nonumber\\
\vec s&=& \vesg-\vesqa;
\label{3160b}
\eeqn
\beqn
\vec R^{\,+}&=& {1\over2}\left(\vec R+\vec S\right);\ \ \ \ \
\vec R^{\,-}= \vec R-\vec S;
\nonumber\\
\vec r^{\,+}&=& {1\over2}\left(\vec r+\vec s\right);\ \ \ \ \
\vec r^{\,-}= \vec r-\vec s.
\label{3170b}
\eeqn
Then the Lagrangian $L_2$ gets the form,
\beqn
L_2 &=&
E\, \vec V^{\,+}\vec V^{\,-}+
x(1-x)\,E\,v^{\,+}\vec v^{\,-}+
{1\over2}\mathcal{E}\,\vec\omega^{\,2}
\nonumber\\ &-&
-{1\over2}\biggl[
a\rho^2+ b \left(\vec r^{\,-}\right)^2+
c \left(\vec R^{\,-}\right)^2 +
2 d\, \vec\rho \vec r^{\,-}
\nonumber\\ &+&
2 f\, \vec\rho \vec R^{\,-}+
2 e\, \vec r^{\,-} \vec R^{\,-}
\biggr],
\label{3200b}
\eeqn
where the coefficients $a,...,f$ were defined in (\ref{1870b}).

In this case the equation of motion has the form,
\beqn
\left(\frac{\partial}{\partial z}\right)^2\vec R^{\,-} &=&
\left(\frac{\partial}{\partial z}\right)^2\vec r^{\,-}=0;
\nonumber\\
\mathcal{E}\,\left(\frac{\partial}{\partial z}\right)^2\vec \tau^{\,-} &+&
a\,\vec\tau=0,
\label{3300b}
\eeqn
where
\beq
\vec\tau=\vec\rho+\frac{\tilde d}{\tilde a}\,\vec r^{\,-}+
\frac{\tilde e}{\tilde a}\,\vec R^{\,-}.
\label{3400b}
\eeq

Similar to other terms in the Lagrangian which were calculated in previous
sections, the $z$-dependence of the functions $\vec R^{\,+}(z)$ and $\vec
r^{\,+}(z)$ does not affect the corresponding action. Using the
solutions of equations (\ref{3300b}) in the Lagrangian
Eq.~(\ref{3200b}) we obtain the action,
\begin{widetext}

\beqn
S_2&=& \int\limits_{z_2}^{z_3} dz\,L_2(z)=
\frac{E}{\Delta z_{23}}\,\biggl\{
\left(\vec R_2^{\,+}-\vec R_1^{\,+}\right)
\left(\vec R_2^{\,-}-\vec R_1^{\,-}\right)+
x(1-x)\left(\vec r_2^{\,+}-\vec r_1^{\,+}\right)
\left(\vec r_2^{\,-}-\vec r_1^{\,-}\right)
\nonumber\\ &+&
{1\over2}\mathcal{E}\omega_0
\biggl[\bigl(\vec\tau_2^{\,2}+\vec\tau_1^{\,2}\bigl)\cot(\omega_0z)-
2\frac{\vec\tau_1\vec\tau_2}{\sin(\omega_0z)}\biggr]+
\frac{\mathcal{E}}{2\Delta z_{23}}\biggl[\left(\vec\rho_1-\vec\rho_2\right)^2-
\left(\vec\tau_1-\vec\tau_2\right)^2\biggr]
\nonumber\\ &-&
{1\over6}\left(b-{d^2\over a}\right)
\left[\bigl(\vec r_2^{\,-}\bigr)^2+
\vec r_2^{\,-}\cdot \vec r_1^{\,-}+
\bigl(\vec r_1^{\,-}\bigr)^2\right]-
{1\over6}\left(c-{f^2\over a}\right)
\left[\bigl(\vec R_2^{\,-}\bigr)^2+
\vec R_2^{\,-}\cdot \vec R_1^{\,-}+
\bigl(\vec R_1^{\,-}\bigr)^2\right]
\nonumber\\ &-&
{1\over6}\left(e-{df\over a}\right)
\left[
2\vec R_2^{\,-}\cdot \vec r_2^{\,-}+
2\vec R_1^{\,-}\cdot \vec r_1^{\,-}+
\vec R_2^{\,-}\cdot \vec r_1^{\,-}+
\vec R_1^{\,-}\cdot \vec r_2^{\,-}
\right],
\label{3500b}
\eeqn
\end{widetext}
where $\omega_0=\sqrt{a/\mathcal{E}}$; $\Delta z_{23}=z_3-z_2$.

Eventually, we get,
\beq
W_2(z)=N_2(\Delta z_{23})\,e^{iS_2},
\label{3600b}
\eeq
where
\beq
N_2(\Delta z_{23})=\frac{-i\mathcal{E}\omega_0}{2\pi\,\sin(\omega_0\Delta z_{23})}\,
\frac{(E\mathcal{E})^2}{(2\pi\Delta z_{23})^4}.
\label{3700b}
\eeq


\end{document}